\documentclass[%
 reprint,
 amsmath,amssymb,
 aps,
 prd,
]{revtex4-2}

\usepackage{geometry}                
\usepackage{graphicx}
\usepackage{amssymb}
\usepackage{amsmath}
\usepackage{epstopdf}
\usepackage{hyperref}
\usepackage{xcolor}
\usepackage{siunitx}
\usepackage{lineno}

\newcommand{\mcp}[1]{{\small\color{red} \bf{[MCP: #1]}}}
\newcommand{\bs}[1]{{\small\color{blue} \bf{[BS: #1]}}}
\newcommand{\xl}[1]{{\small\color{green} \bf{[XL: #1]}}}

\newcommand{\DisableComments}[0]{
\renewcommand{\mcp}[1]{}
\renewcommand{\bs}[1]{}
\renewcommand{\xl}[1]{}
}
\DisableComments

\begin{document}

\title{Modeling the Differential Rate for Signal Interactions in Coincidence with Noise Fluctuations or Large Rate Backgrounds}
\author{Xinran Li}
\email{xinranli@lbl.gov}
\affiliation{%
 Lawrence Berkeley National Laboratory, Berkeley, CA 94720, USA
}
\author{Matt Pyle}
\affiliation{%
 Lawrence Berkeley National Laboratory
}
\affiliation{%
 University of California Berkeley, Berkeley, CA 94720, USA
}
\author{Bernard Sadoulet}
\affiliation{%
 University of California Berkeley, Berkeley, CA 94720, USA
}
\date{\today}                                           

\begin{abstract}
    The characteristic energy of a relic dark matter interaction with a detector scales strongly with the putative dark matter mass. Consequently, experimental search sensitivity at the lightest masses will always come from interactions whose size is similar to noise fluctuations and low energy backgrounds in the detector.

    In this paper, we correctly calculate the net change in measured differential event rate due to low rate signal interactions that overlap in time with noise and backgrounds, accounting for both periods of time when the signal is coincident with noise/backgrounds and for the decreased amount of time in which only noise/backgrounds occur and we show that the introduction of  random simulated signal events into the continuous raw data stream (a form of ``salting") provides a correct and practical implementation to estimate this net linear signal sensitivity. 

    Unfortunately this does not apply to the situation with non negligible pile-ups as we show through explicit examples. Consequently, this exclusion technique should be complemented by less sensitive techniques that can separately exclude large rate, high pileup signal parameter space. An analysis threshold above the Gaussian-like noise component of the measured differential rate spectrum should also likely produce conservative limits. 

    Though previous light mass dark matter searches did not correctly account for decreased background only live time effects in their signal sensitivity, conservative analysis choices likely protected their published limits from being non-conservative. 
\end{abstract}

\maketitle

\section{Introduction}
\label{sec:introduction}

Detectors for direct detection of particle dark matter (DM) record signals in time streams. In most cases, a triggering algorithm is applied to select potentially interesting signals from the continuous stream. Then, the signals are processed to estimate the differential rate as a function of measured energy, $\frac{dR}{dE'}$, and generate a DM sensitivity estimate.  
Detectors have noise and backgrounds, and an event with true energy deposition $E$ produces a signal reconstructed at energy $E'$ with a probability $f(E'|E)$. Understanding $f(E'|E)$, i.e., the detector energy response, is the core of an experiment. 

For light mass DM, where the mass of the DM, $M_\mathrm{DM}$, is much less than the mass of the target nuclei in the detector, $M_\mathrm{N}$, the energy deposited in an elastic two-body nuclear scattering interaction is $E \lesssim \frac{2 M_\mathrm{DM}^{2}v_\mathrm{esc}^{2}}{M_\mathrm{N}}$, where $v_\mathrm{esc}$ is the escape velocity for the Milkyway. This $M_\mathrm{DM}^{2}$ dependence means that signals for the lowest $M_\mathrm{DM}$ are near or even below the trigger threshold, $E_{T}'$. Specifically, it is possible that a DM interaction with a true energy deposition below the trigger threshold ($E<E'_{T}$) could occur in coincidence with a positive random noise fluctuation, $\delta E'$, and thus be boosted above the trigger threshold, $E+\delta E'= E' > E'_{T}$. We want to correctly derive the change in the measured differential rate spectrum, $\frac{dR}{dE'}$, due to these interactions so that we can correctly account for this sensitivity in our searches.

As we will elaborate in section \ref{sec:digitized device}, the net change in $\frac{dR}{dE'}$ is due to both:
\begin{itemize}
    \item the signal interacting with the detector in coincidence with a large noise fluctuation or other background. 
    \item a decrease in the rate of large noise fluctuations/backgrounds that are non-coincident with signal interactions. 
\end{itemize}
Not accounting for this latter effect can lead to overestimation of the experimental sensitivity. For example, an experiment must obviously have no sensitivity to a hypothetical interaction that deposits identically zero true energy in the detector. However, this $E=0$ event will have a probability of being in coincidence with background fluctuations $\delta E' > E'_{t}$ and thus will naively be boosted above the trigger threshold. Only after accounting for both of the effects above will $\frac{dR}{dE'}$ not have unphysical sensitivity to $E=0$ interactions (Sec.~\ref{sec:MDMto0}).

Though we were motivated to understand noise boosting originally in phonon detectors, this is a general problem present in all detector technologies and in any search where the signal is commingled, coincident, and of similar magnitude to the noise and background events. For example, 
a light mass DM search with a p-type point contact germanium ionization detector \cite{liu2022studies} attempted to search for noise-boosted signals by simply Gaussian smearing the expected DM spectrum without correctly restricting the boost and thus potentially overestimated their DM sensitivity for light mass DM. 

We will first examine a simple scenario of ideal integrating detectors, and introduce the concept of the net differential signal response $\Delta f(E'|E) =f(E'|E)-f(E'|0)$ in Sec.~\ref{sec:digitized device}. In Sec.~\ref{sec:noDM}, we will discuss the subtlety of estimating $\Delta f(E'|E)$ when the experimental search data is contaminated with an unknown rate of signal interactions and find that as long as the signal pileup rate is negligible, a conservative interaction limit that is guaranteed to have a larger expectation than the true signal rate can be estimated using the net differential response. Unfortunately, we also show that for a true signal with large pileup, limits are not guaranteed to be conservative. Thus, limits based on the net linear differential sensitivity should ideally be complemented by other limit setting techniques which may be less sensitive but are guaranteed to be conservative in the pileup limit.

In Sec.~\ref{sec:digitized device generic}, we will generalize the discussion to continuous DM interaction spectra and demonstrate how conservativeness depends on true signal pileup in a toy example. With this example as a guide, we propose that setting limits based only upon measured energies that are above the approximately Gaussian noise blob will likely produce conservative limits, even though this isn't strictly proven. We will also discuss the likely conservativeness of analysis strategies used in previous light mass DM direct detection searches 
\cite{CPDv1_PRL21_LDMSearch,CRESST3_PRD19_FirstDMLimits,CRESST_PRD24_SOS_DMLimits,EDELWEISS_PRD19_LDMSearch}. All of the above has been done with an ideal integrating detector that is easy to understand. Thus, we will generalize these concepts to experiments with continuous data streams that have more complicated event timing information in Sec. \ref{sec:waveform_detector}. We will refer to these detectors as ``waveform detectors". In such waveform devices, the concept of a signal interaction being in ``coincidence" with a large noise fluctuation is qualitative. More rigorously, we should say that the time difference between a large noise fluctuation and the true signal event is small compared to the pulse width. Finally, in Sec. \ref{sec:salting}, we will show that introducing a known rate of artificial random signal events in the continuous raw data stream (pretrigger ``salting") correctly estimates the net differential response.
\section{Derivation for a Discretized Integrating Device} 
\label{sec:digitized device}

In this section, we will suppress these subtleties and think instead about the behavior of an idealized integrating time-discretized detector like a CCD.
In an idealized CCD, for instance, the total integrated charge deposited in the pixel is measured every $\Delta t$, after which the pixel is reset and begins to integrate the charge for the next discrete measurement. All measurements are recorded without a trigger. We will assume that the time scale for energy deposition is instantaneous and the time to measure and reset is infinitesimal. 

\subsection{Single Dirac-Delta True Energy Deposition} 
\label{sec:digitized device Bosonic}

The second simplification we will make, initially at least, is that each and every interaction event produces an identical true energy signal in the sensor, $E_{s}$. Beyond being easier to understand, this scenario has physical relevance since it corresponds to Bosonic DM absorption.

With these two simplifications, we can calculate the expected differential rate of the measured energy $E'$ in terms of:
\begin{itemize}
    \item $f(E' | E= n_{s} E_{s})\equiv f(E' | n_{s} E_{s})$ : the probability distribution of measured energy, $E'$, in the time interval $\Delta t$ given a true deposited energy of $E= n_{s} E_{s}$ from  $n_{s}$ signal interactions occurring in the same time interval.
    \item $f(E' | E= 0) \equiv f(E'|0)$  : the probability distribution of measured energy, $E'$, in the time interval $\Delta t$ with no signal interactions in the same time interval. This is a subgroup of the previous definition. We emphasize that this is everything that is not due to the signal. It potentially includes backgrounds interactions, calibration source interactions, all convolved with various noise sources. 
    \item $P(n_{s}| \lambda_{s}=R_{s}\Delta t)$ : the Poissonian distribution for having $n_{s}$ events within a time $\Delta t$ with an average number of interactions $\lambda_{s}$ or equivalently an interaction rate of $R_{s}$.
\end{itemize}			

Please note a few things. First, $f(E' | n_{s} E_{s})$ and $f(E'|0)$ are highly related.  $f(E'|n_sE_s)$ can be directly measured by explicitly \textbf{salting} no-signal measurements that are pulled from $f(E'|0)$. Explicitly, one would simply deposit $n_s$ signals with energy $E_s$ in each and every measurement period, $\Delta t$. 
More simply, for a perfectly linearly idealized integrating device where the energies of all the events that occur within the $\Delta t$ time bin just simply add
\begin{equation}
f(E' | n_{s} E_{s})= f(E'-n_{s} E_{s}| 0)
\label{eq:f(E'|Es)fromf(E|0)}
\end{equation}

Using these distributions, we can then write the expected signal rate $\frac{dR}{dE'}$ as 
\begin{equation}
\begin{split}
\frac{dR}{dE'}(E'|S(E_{s},R_{s})) = \frac{1}{\Delta t} f(E'|S(E_{s},R_{s}))& \\
 = \frac{1}{\Delta t}\sum_{n_{s}=0}^{\infty} P(n_{s}| R_{s} \Delta t) f(E' | n_{s} E_{s})&
\label{eq:dRdE'_full}
\end{split}
\end{equation}
We use $S(E_{s},R_{s})$ to represent the signal model. For simplicity, in the later sections, the parameters for the signal model in context will only be explicitly written out once, and then the model will be noted as $S$. This notation allows us to easily generalize to more complex signal models, which can produce a continuum of true energy depositions, like DM scattering (see Sec.~\ref{sec:digitized device generic}).

We explicitly highlight the seemingly obvious relationship between the measured differential rate and the probability distribution of the measured energy, $\frac{dR}{dE'}(E'|S)=\frac{1}{\Delta t}f(E'|S)$. In particular, this relationship means that ${\Delta t}\int dE' \frac{dR}{dE'}(E'|S)=1$. By contrast, the integral of the differential rate of the signal with respect to the true energy deposition, $\frac{dR}{dE}(E|S)$, is ${\Delta t}\int dE \frac{dR}{dE}(E|S)=\lambda$, and is thus \textbf{not} a probability distribution. To gain a deeper understanding for this difference, we look at the two limiting cases. When there is no signal interaction at all, ${\Delta t}\int dE \frac{dR}{dE}(E|S)=\lambda_{s}=0$. However, the measured differential rate is clearly non-zero; it's a measure of the noise and backgrounds. On the other hand, when $\lambda\gg1$, there is significant pileup of multiple signal interactions in every bin and $\frac{dR}{dE'}$ bears little resemblance to $\frac{dR}{dE}$. 

Returning to Eq.~\ref{eq:dRdE'_full}, we split off and rewrite the $n_{s}=0$ term using the fact that $\sum_{n_{s}=0}^{\infty} P(n_{s})=1$, and find
\begin{widetext}
\begin{equation}
\begin{split}
\frac{dR}{dE'}(E'|S)&= \frac{1}{\Delta t} f(E' |0)\left(1-\sum_{n_{s}=1}^{\infty} P(n_{s}| R_{s} \Delta t) \right)+ \frac{1}{\Delta t} \sum_{n_{s}=1}^{\infty} P(n_{s}| R_{s} \Delta t) f(E' | n_{s} E_{s}) \\
&=  \frac{dR}{dE'}(E'|0) \left({1-\sum_{n_{s}=1}^{\infty} P(n_{s}| R_{s} \Delta t) }\right)+
        \frac{1}{\Delta t} \sum_{n_{s}=1}^{\infty} P(n_{s}| R_{s} \Delta t) f(E' | n_{s} E_{s})\\
\end{split}
\label{eq:dRdE'_full2}
\end{equation}
\end{widetext}
since $\frac{dR}{dE'}(E'|0) = \frac{1}{\Delta t} f(E' | 0)$.

Eq.~\ref{eq:dRdE'_full2} shows that signal interactions affect $\frac{dR}{dE'}$ in two distinct ways, \textbf{both of which must be accounted for}. The second term, $ \frac{1}{\Delta t} \sum_{n_{s}=1}^{\infty} P(n_{s}| R_{s} \Delta t) f(E' | n_{s} E_{s})$, accounts for the change in the differential rate distribution from those time periods which have one or more signal interactions. The $\frac{dR}{dE'}(E'|0) \left({1-\sum_{n_{s}=1}^{\infty} P(n_{s}| R_{s} \Delta t) }\right)$ term, by contrast, accounts for the fact that as $R_{s}$ increases, there are fewer time periods without a signal interaction; there are fewer time periods that are solely noise. In other words, the presence of signal interactions decreases the live time available for pure noise fluctuations to be recorded.

Grouping terms in Eq.~\ref{eq:dRdE'_full2} together by $P(n_{s}|R_s \Delta t)$, we find that 
\begin{equation}
\begin{split}
&\frac{dR}{dE'}(E'|S) =  \frac{dR}{dE'}(E'|0) \\
&+ \frac{1}{\Delta t} \sum_{n_{s}=1}^{\infty} P(n_{s}| R_{s} \Delta t) \left({ f(E' | n_{s} E_{s})-f(E' | 0)} \right)
\end{split}
\label{eq:dRdE'_full3}
\end{equation}
or equivalently by multiplying through by $\Delta t$, we estimate the probability distribution for the measured $E'$ in each integration period:
\begin{equation}
\begin{split}
&f(E'|S) = f(E'|0) \\
&+ \sum_{n_{s}=1}^{\infty} P(n_{s}| R_{s} \Delta t) \left({ f(E' | n_{s} E_{s})-f(E' | 0)} \right)
\end{split}
\label{eq:f(E'|S)_full3}
\end{equation}
Under the additional assumption that there is minimal signal pileup, $R_{s}\Delta t \ll 1$, the dominant term in the sum is $n_{s} = 1$ and this simplifies drastically to
\begin{equation}
\begin{split}
&\frac{dR}{dE'}(E'|S)  \\
&= \frac{dR}{dE'}(E'|0) +  R_{s}( f(E' | E_{s})- f(E' | 0))
\end{split}
\label{eq:dRdE'_linear}
\end{equation}
or equivalently
\begin{equation}
\begin{split}
&f(E'|S) \\
&= f(E'|0) +  R_{s}\Delta t( f(E' | E_{s})- f(E' | 0))
\end{split}
\label{eq:f(E'|S)'_linear}
\end{equation}

When written in this way, the term $R_{s}( f(E' | E_{s})- f(E' | 0))$ should be thought of as the net change of the differential rate due to a signal interaction with rate $R_{s}$ and energy $E_{s}$. We define 
\begin{equation}
    \Delta f(E'|E_{s}) \equiv f(E' | E_{s})- f(E' | 0)
\label{eq:NetDiffSens}
\end{equation}
as the ``net differential response'' of the detector to an event of true energy $E_{s}$. 

Eq. \ref{eq:dRdE'_linear} (and its non-linear generalization in Eq. \ref{eq:dRdE'_full3} that accounts for signal pileup) is the core insight of this paper. It describes how a signal that is overlapping with background and noise affects the experimentally measured differential rate, $\frac{dR}{dE'}(E'|S)$. Most importantly, correct modeling of the $\Delta f(E'|E_{s})$ must occur no matter the statistical technique ultimately used to estimate experimental search sensitivity.

\subsection{A Variety of Test Cases}
We will discuss a series of intuitive test cases to demonstrate that Eq.~\ref{eq:dRdE'_linear} and their DM high rate generalizations (Eq.~\ref{eq:dRdE'_full3}) should be used rather than expressions which don't account for the decrease in background only time periods (Eq.~\ref{eq:dRbaddE'_bosonic}) in any analysis where there is a significant noise/background event rate that overlaps with the signal. 
 
\subsubsection{Bosonic DM search with a Gaussian random noise}
\label{sec:Discritized_SimpleGaussianNoise}
  \begin{figure}[h!]
    \centering
     \includegraphics[width=\linewidth]{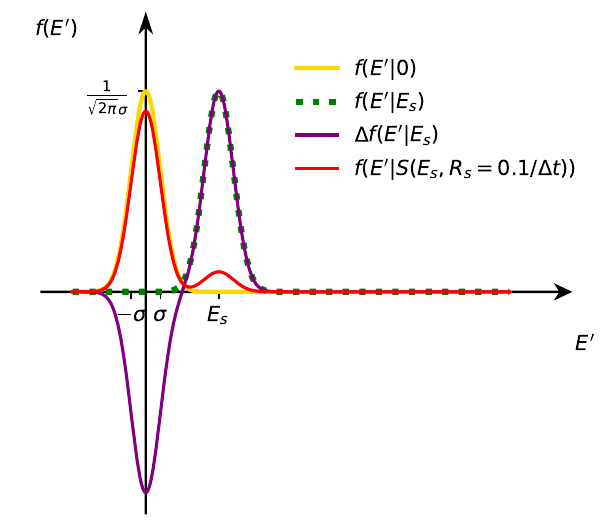} 
     \caption{Probability distribution functions for when $f(E', 0)$ (yellow) is Gaussian distributed noise.The net change $\Delta f(E'|E_s)$ (purple) is $f(E'|E_{s})$ (dashed green) subtracted by $f(E', 0)$. The DM signal energy $E_{s}=5\sigma$, where $\sigma$ is the Gaussian RMS. The signal rate $R_{s}=0.1/\Delta t$. }
\label{fig:f_gaussian}
\end{figure}
We will start with the ideal scenario, a detector for Bosonic DM absorption with Gaussian noise fluctuations. In Fig. \ref{fig:f_gaussian}, there are 2 peaks visible in $f(E' |S(E_s=5\sigma,R_s=0.1/\Delta t))$ (red). The large peak centered at $E'=0$ consists of time frames that are not in coincidence with a signal interaction; they are due to noise-only time frames. The peak centered at $E'=E_{s}$ consists of time frames that have a DM interaction in coincident with the noise. 

Notice that the rate of events around zero is slightly less for $f(E'|S)$ than $f(E'|0)$ because the former is only those noise events that are anti-coincident with the DM. Those coincident with the DM have been up-shifted to being centered around $E_{s}$. This is a physical effect represented by the $f(E'|0) \left({1-\sum_{n_{s}=1}^{\infty} P(n_{s}| R_{s} \Delta t) }\right)$ Eq. \ref{eq:dRdE'_full2} or equivalently the $-R_{s}\Delta t f(E'|0)$ term in Eq. \ref{eq:f(E'|S)'_linear}. 

In experiments that search for signals which are well above the noise peak, like LZ \cite{aalbers2023firstLZ}, CDMS-II \cite{agnese2013siliconCDMSII}, SuperCDMS Soudan  \cite{agnese2018firstSuperCDMS}, EDELWEISS III  \cite{EDELWEISS3_EPJC16_WIMPDMSearch}, CRESST III \cite{CRESSTIII:EPJC16:WIMPSearch} etc., it's reasonable to not include $f(E'|0)$ term since there is no or very minimal overlap with $f(E'|E_{s})$. Thus, there exists an analysis threshold that is above nearly the entirety of $f(E'|0)$ while being below the majority of $f(E'|E_{s})$. Consequently, neglecting the decrease in the noise-only event rate has no scientific significance in those experiments. By contrast, for light mass DM searches where the signals are small, $f(E'|E_{s})$ can have significant overlap with $f(E'|0)$, and thus one must use the net differential sensitivity.

\subsubsection{Searches for Dark Matter where \texorpdfstring{$M_\mathrm{DM}\rightarrow 0$}{M DM approaches 0}}
\label{sec:MDMto0}
As the mass of dark matter, $M_\mathrm{DM}$, approaches 0, the maximum possible energy transferred to the detector also approaches zero, $E_{s} \rightarrow 0$. As such, a necessary but certainly not sufficient condition for a valid DM search analysis technique is that it has no sensitivity to DM interactions in the limit as $M_\mathrm{DM} \rightarrow 0$.
Notice that Eq.~\ref{eq:f(E'|S)'_linear} naturally satisfies this condition
 \begin{equation}
 \begin{split}
 &f(E'|S(E_{s}\rightarrow 0,R_{s})) \\
 &= f(E'|0) +  R_{s}\Delta t( f(E' | E_{s} \rightarrow 0)- f(E' | 0))\\
 &= f(E'|0) - R_{s}\Delta t \lim_{E_{s}\rightarrow 0}\frac{df(E'|0)}{dE'}E_{s}+... \\
 &= f(E'|0)
\end{split}
\label{eq:df_0Es}
\end{equation}
In this limit, the measured distribution $f(E'|S)$ is by construction independent of $R_{s}$, and thus, the experiment has absolutely no sensitivity to DM. By contrast, \cite{EDELWEISS_PRD19_LDMSearch, CPDv1_PRL21_LDMSearch, CRESST3_PRD19_FirstDMLimits} 
did not directly account for decreased noise only live time and incorrectly estimated their sensitivity was
\begin{align*}
 &\frac{dR_\mathrm{NoLT}}{dE'}(E'|S(E_{s},R_{s})) \\
 &=  \frac{dR}{dE'}(E'|0) +  R_{s} f(E' | E_{s})
\end{align*}
or expressed by the measurement probability distribution as
\begin{equation}
\begin{split}
   & f_\mathrm{NoLT}(E'|S(E_{s},R_{s})) \\
   &  = f(E'|0) +  R_{s}\Delta t f(E' | E_{s})
\end{split}
    \label{eq:dRbaddE'_bosonic}
\end{equation} 
where ``NoLT'' stands for no live-time correction. This expression is no longer a proper probability distribution, since the total probability is no longer unity. Additionally, in the limit of $E_{s} \rightarrow 0$, this incorrect formulation calculates
 \begin{align*}
 &f_{\mathrm{NoLT}}(E'|S(E_{s}\rightarrow 0,R_{s})) \\
 &=  f(E'|0) +  R_{s} \Delta tf(E' | E_{s} \rightarrow 0)\\
 &=  f(E'|0) (1+R_{s}\Delta t)
\end{align*}
a dependence of the differential rate on $R_{s}$. This is clearly unphysical. It falsely claims that the number of high-energy noise events increases with the DM interaction rate, even though these interactions deposit zero energy, and thus each of the historical analyses that used Eq.~\ref{eq:dRbaddE'_bosonic} (or more precisely its generalized form for continuum signals found in Eq.~\ref{eq:dRbaddE'}) placed additional artificial constraints on their analyses to mitigate these issues as discussed in Sec.~\ref{sec:PrevExpDiscussion}
  
\subsubsection{Effect of slowly varying non-Gaussian noise tails}
 \label{sec:DMsearch_flattail}
Next, we consider the scenario where the no-signal distribution has a non-Gaussian, monotonically decreasing tail of unknown origin. This scenario is extremely pertinent since the current generation of light mass DM searches based on cryogenic calorimeters measures a poorly understood near threshold background excess of this type\cite{CPDv1_PRL21_LDMSearch, hehn2016improved, adari2022excess}. 

\begin{figure}
    \centering
     \includegraphics[width=\linewidth]{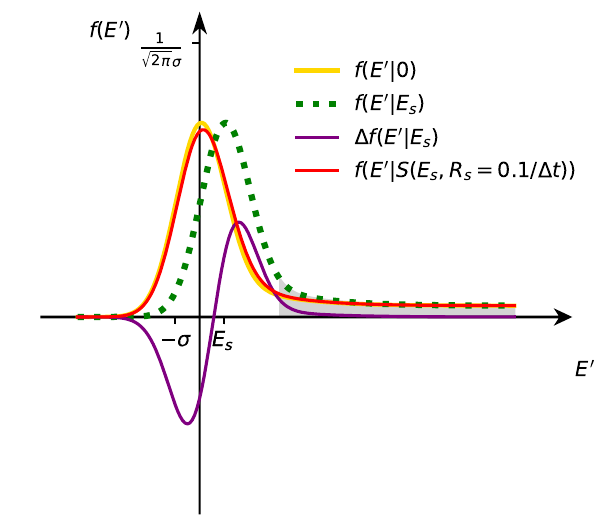} 
    \caption{Same probability distributions as Fig.~\ref{fig:f_gaussian} but assuming a non-Gaussian noise. The signal model here represents a case with low DM energy, where $E_s=\sigma$ and $R_s=0.1/\Delta t$. Gray highlights the non-Gaussian noise and background region where $\Delta f(E'|E_s)$ is suppressed compared to $f(E'|E_s)$. The flat background differential rate is $0.02/\Delta t/\sigma$.}
\label{fig:f_flattail}
\end{figure}

For concreteness, we model this scenario with an exponential plus a flat background, which qualitatively accounts for the low energy excess backgrounds and environmental radioactive backgrounds, that has been smeared by Gaussian noise (Fig. \ref{fig:f_flattail}).

In the tail region (gray in Fig.~\ref{fig:f_flattail}) of the noise spectrum, the net sensitivity response (purple in Fig.~\ref{fig:f_flattail}) is found to be significantly suppressed compared to $f(E'|E_{s})$(dashed green). This is due to the fact that for small DM energy depositions, $E_{s}$, $\Delta f(E'|E_s) \sim -\frac{df(E'|0)}{dE'}E_{s}+ ...$(Eq. \ref{eq:df_0Es}); it's proportional to the slope of $f(E'|0)$. \textbf{Thus, coincidence with a perfectly flat background tail can not be used to gain sensitivity to subthreshold signals.} Again, the historically used, but incorrect, rate model of Eq.~\ref{eq:dRbaddE'} overestimates signal sensitivity in this critical noise scenario. This example,  strongly motivates the use of the correct net differential signal response function in future experiments with such backgrounds so that the actual sensitivity can be correctly estimated. 

\section{Estimating the signal rate}
 \label{sec:noDM}
 
Now that we understand theoretically how $f(E')$, and equivalently $\frac{dR}{dE'}$, varies with a known signal rate $R_{s}$ of true energy depositions $E_{s}$ in the linear (Eq. \ref{eq:dRdE'_linear}) and non-linear regimes (Eq. \ref{eq:dRdE'_full3}), we can attempt to statistically estimate an unknown measured signal rate in experimental search data.

The first step, of course, is to simply measure the potentially signal contaminated differential rate, $\widehat{\frac{dR}{dE'}}(E'|S(E_{s},\lambda_{s}=R_s\Delta t)))$, where the $S$ explicitly highlights that the measured search data is potentially contaminated by an unknown signal with rate $R_{s}$ and true energy deposition $E_{s}$ while the ` $\hat{}$ ' communicates that this is a statistical estimator. Specifically, we can count the number of measurements, $N_{E'}$, with an estimated energy $E'$ within the range of $[E'-\frac{\Delta E'}{2}, E'+\frac{\Delta E}{2} ]$ over a time $T_\mathrm{total}$. In the limit of small energy intervals, the estimator becomes
\begin{equation}
\begin{split}
&\widehat{\frac{dR}{dE'}}(E'|S) \\
&\equiv \lim_{\Delta E' \rightarrow 0} \frac{N_{E'}([E'-\frac{\Delta E'}{2}, E'+\frac{\Delta E}{2} ])}{\Delta E' T_\mathrm{total}}
\end{split}
\end{equation}

This estimator and the related estimator for $\hat{f}(E'|S)$ are both unbiased and consistent with the true value \cite{Kendall_Stuart_Estimation} since their expectation matches the true value and therefore 
\begin{widetext}
\begin{equation}
 \langle \widehat{\frac{dR}{dE'}}(E'|S)\rangle \Delta t = \langle \hat{f}(E'|S)\rangle 
= f(E'|0) + \sum_{n_{s}=1}^{\infty} P(n_{s}|\lambda_{s}) \left({ f(E'| n_{s} E_{s})-f(E'| 0)} \right)
\label{eq:dRdE'_bkgsub}
\end{equation}
\end{widetext}
where the second equality follows from Eq.~\ref{eq:f(E'|S)_full3}. To generate an unbiased and consistent estimator of $\lambda_{s}=R_{s}\Delta t$ from Eq.~\ref{eq:dRdE'_bkgsub} clearly requires an additional estimation of background + noise distribution \textbf{without any potential signal contamination}, $f(E'|0)$. If this is possible, then $\hat{f}(E'|n_{s}E_{s})$ can be generated either by shifting the distribution in the case of ideal linear integrating detectors, $\hat{f}(E'|n_{s}E_{s}) = \hat{f}(E'- n_{s}E_{s}|0)$ or via salting methods (Eq.~\ref{eq:f(E'|Es)fromf(E|0)}). With this additional knowledge, $\hat{\lambda_{s}}(E')$ is the solution of
\begin{equation}
\begin{split}
  &\hat{f}(E'|S) = \hat{f}(E'|0) \\
  &+ \sum_{n_{s}=1}^{\infty} P(n_{s}|\hat{\lambda_{s}}) \left({ \hat{f}(E'| n_{s} E_{s})-\hat{f}(E'| 0)} \right)
\end{split}
\label{dRdE'_bkgsub2}
\end{equation}
which in the small pileup limit becomes
\begin{equation}
    \hat{R_s}(E'|E_s) \Delta t= \hat{\lambda}_{s}(E')=\frac{\hat{f}(E'|S)-\hat{f}(E'|0)}{ \hat{f}(E' | E_{s})- \hat{f}(E' | 0)}
\label{eq:dRdE'_bkgsub_linear}
\end{equation}
where we've explicitly highlighted that $\hat{R_s}(E'|E_s)$ is a function of $E'$. 
 More complex estimators can be constructed by an optimal weighted integration of $\hat{R_{s}}$ over a range of $E'$ \cite{yellin2002finding} or even using maximum likelihood methods, but they basically share the same features as Eqs. \ref{dRdE'_bkgsub2} and \ref{eq:dRdE'_bkgsub_linear}.

\subsection{The difficulties of measuring \texorpdfstring{$f(E'|0)$}{f(E'|0)}}
\label{sec:difficulty_f0}

Unfortunately, in a large class of experiments, it is impossible to ``turn off" the DM interaction signal to directly measure $f(E'|0)$. High mass DM experiments like the noble gas two phase TPCs, for example,  depend upon a wealth of event information (ionization and scintillation signal amplitudes, position information) and an enormous amount of calibration data to both separate the majority of backgrounds and model the residual overlapping background $f(E'|0)$. Unfortunately, the current generation of light DM matter experiments have minimally understood indistinguishable backgrounds (charge leakage \cite{aguilar2024confirmation, albakry2022investigating}, dark counts \cite{akerib2020investigationLUXSingleElectron,agnes2018low}, low energy event excess \cite{anthony2024stress, CPDv1_PRL21_LDMSearch, CRESST3_PRD19_FirstDMLimits,CRESST_PRD24_SOS_DMLimits}) and thus an understanding of $f(E'|0)$ is currently impossible. 

\subsection{Measuring the net differential response}
\label{sec:netdiffresp}
\bs{small reorganization with new subsection to emphasize that the net differential response is measurable in the no pileup regime }
While in many cases $f(E'|0)$ is not directly estimable and thus a background subtracted signal rate is impossible, one can still estimate the net differential response $\Delta f(E'|E_{s}) \equiv f(E' | E_{s})- f(E' | 0)$ (Eq.~\ref{eq:NetDiffSens}) from potentially contaminated search data in \textbf{the no pileup limit}.

To prove this let us take the probability distribution, $f(E'|S)$, that is potentially contaminated by the signal  $S(E_s,\lambda_s)$, and generate the distribution $f(E'|S+n_{s}E_{s})$ either by shifting the distribution in the case of ideal linear integrating detectors, $f(E'|S+n_{s}E_{s}) = f(E'- n_{s}E_{s}|S)$ or by explicitly adding a true energy deposition of $n_{s}E_{s}$ to the search data before measurement via salting (Eq.~\ref{eq:f(E'|Es)fromf(E|0)}).
Following from Eq.~\ref{eq:dRdE'_bkgsub}, we can relate these boosted signal contaminated search distributions to that of the boosted background-only distributions: 
\begin{widetext}
\begin{equation}
f(E'|S+n_{s}E_{s}) = f(E'|n_{s}E_{s}) + \sum_{m_{s}=1}^{\infty} P(m_{s}|\lambda_{s}) \left({ f(E'| (m_{s}+n_{s})E_{s})-f(E'| n_{s}E_{s})} \right)
\label{eq:fS?(E'|S?+nEs)}
\end{equation}

Flipping Eqs.~\ref{eq:dRdE'_bkgsub} and \ref{eq:fS?(E'|S?+nEs)} to write $f(E'|0)$ and $f(E'|n_{s}E_{s})$ in terms of these measurable search distributions and a higher order signal pileup residuals, we find
\begin{equation}
f(E'|0)= f(E'|S) - \sum_{m_{s}=1}^{\infty} P(m_{s}|\lambda_{s}) \left({ f(E'| m_{s} E_{s})-f(E'|0)} \right)
\label{eq:f(E'|0)_FromSearch)}
\end{equation}
and
\begin{equation}
f(E'|n_s E_{s})= f(E'|S+n_s E_{s})
-\sum_{m_{s}=1}^{\infty}\! P(m_{s}|\lambda_{s}) \left({ f(E'|(n_{s}\!+\!m_{s})\! E_{s})-f(E'|n_s E_{s})} \right)
\label{eq:f(E'|nsEs)_FromSearch}
\end{equation}
which we can then use to rewrite the unmeasurable net differential sensitivities in Eq.~\ref{eq:dRdE'_bkgsub} in terms of measurable differences between signal-contaminated distributions and unmeasurable signal pileup residuals:
\begin{equation}
\begin{split}
f(E'&|S) =f(E'|0)\! + \! \sum_{n_{s}=1}^{\infty} P(n_{s}| \lambda_{s}) \left(f(E'|S+n_{s}\! E_{s})\!-f(E'| S) \right)\\
            &-\sum_{n_{s}=1}^{\infty}\! \sum_{m_{s}=1}^{\infty}\! P(n_{s}|\lambda_{s}) P(m_{s}| \lambda_{s}) \left[f(E'|(m_{s}\!+\!n_{s})E_{s})\!-\!f(E'|n_{s}E_{s})\!-\!f(E'|m_{s}E_{s}) \!+\! f(E'|0)\right]
\end{split}
\label{eq:dRdE'_conservative3}
\end{equation}

If the unknown average signal interaction number in each time bin, $ \lambda_{s}= R_{s} \Delta t$, is \textbf{already} known to be $\ll 1$, then Eq.~\ref{eq:dRdE'_conservative3} can be Taylor expanded to first order $\lambda_{s}$ and we find
\begin{equation}
\begin{split}
f(E'|S)
&= f(E'|0) + \lambda_{s} \left( f(E'|S+E_{s}) - f(E' |S) \right)\\
    &= f(E'|0) + \lambda_{s} \Delta f(E'|S+E_{s})
\end{split}
\label{eq:dRdE'_conservative_linear1}
\end{equation}
where $\Delta f(E'|S+E_{s})\equiv  f(E'|S+E_{s}) - f(E' |S) $ is the measurable net differential signal sensitivity with potential signal contamination. 
In other words, Eq.~\ref{eq:dRdE'_conservative_linear1} specifically shows that signal contamination doesn't significantly affect the estimate of the net differential sensitivity in the no-pileup regime. 
\clearpage
\end{widetext}

\subsection{Conservative upper bound in the no-pileup regime} 
\label{sec:DMlim_Conservative}
Using the result of the previous section, we can derive  in the no-pileup regime a conservative upper bound on the presence of a signal, $\hat{R}_\mathrm{lim}$, which requires no understanding of the noise, backgrounds and possibly contaminating signals $S$. 

A conservative upper bound, $\hat{R}_\mathrm{lim}(E'|E_{s})$, can be defined as
\begin{equation}
\widehat{R}_\mathrm{lim}\equiv \frac{1}{\Delta t}\frac{\hat{f}(E'|S)}{\widehat{\Delta f}(E'|S+E_{s})} 
\label{eq:dRdE'_conservative}
\end{equation}
whose value for large exposure becomes
\begin{equation}
\begin{split}
\widehat{R}_\mathrm{lim}(E'|E_{s})&\xrightarrow{T_\mathrm{total}\to\infty} \frac{1}{\Delta t}\frac {f(E'|S)}{\Delta f(E'|S+E_{s})}\\
&=\frac{1}{\Delta t}\frac{f(E'|0)}{\Delta f(E'|S+E_{s})} + R_{s}
\label{eq:dRdE'_conservative_linear2}
\end{split}
\end{equation}
Since $f(E'|0)$ is non-negative everywhere, $\lim_{T_\mathrm{total}\rightarrow \infty} \widehat R_\mathrm{lim} \geq R_{s}$ for all $E'$ where $\Delta f(E'|S+E_{s}) > 0$ in the no pileup regime. \textbf{It is conservative for all possible background scenarios in the no-pileup regime}. Of course, more complex upper limits that are constructed by weighted integration of $ \widehat R_\mathrm{lim}(E')$ taking into account the various statistical penalties \cite{yellin2002finding} will also be conservative.

\subsection{Lack of conservativeness in the pileup regime (\texorpdfstring{$\lambda_{s}\gg1$}{lambda s>>1})}
\bs{added lack of}
\label{sec:DMlim_pileup}
In Sec.~\ref{sec:DMlim_Conservative}, we showed that when $\lambda_{s}\ll 1$, $ \widehat R_\mathrm{lim}$ conservatively overestimates $R_{s}$ for all possible background scenarios. We didn't, however, explicitly show that $ \widehat R_\mathrm{lim}$ could be non-conservative and potentially underestimate $R_{s}$ in the pileup regime. Below are two important background scenarios in which $ \widehat R_\mathrm{lim}$ unfortunately underestimates $R_{s}$. They give us significant intuition into scenarios where we can and can not conservatively apply Eq.~\ref{eq:dRdE'_conservative_linear2}.

The absolute most challenging scenario to make conservative is one where the measured search data spectrum is entirely due to dark matter signal interactions plus Gaussian noise, $N(E'|\sigma)$. 

Therefore to calculate $f(E'|S)$ we use Eq.~\ref{eq:dRdE'_full} to allow for the possibility of $m$ signal coincidences within a single time bin:
\begin{equation}
\begin{split}
    f(E'|S) &= \sum_{m=0}^{\infty} P(m| \lambda_{s}) N(E'-m E_{s}|\sigma)\\
    &=\sum_{m=0}^{\infty} \frac{\lambda_{s}^{m} e^{-\lambda_{s}}}{m!} N(E'-m E_{s}|\sigma)
\end{split}    
\label{eq:f?_DM}
\end{equation}

\subsubsection{Non-Conservative Examples: \texorpdfstring{$\lambda_{s}\gg1$}{lambda s >> 1} and Gaussian Noise with \texorpdfstring{$\sigma \ll E_{s}$}{sigma << E s}} 
For the limit where $\sigma \ll E_{s}$, the signal contaminated search data consists of isolated quantized peaks corresponding to $m$ DM coincidences in a single time bin. $f(E'|S+E_{s})$ is found by shifting the $f(E'|S)$ by $E_{s}$
\begin{equation}
\begin{split}
    &f(E'|S + E_{s}) = f(E'-E_{s}|S) \\
    &=\sum_{m=0}^{\infty} \frac{\lambda_{s}^{m} e^{-\lambda_{s}}}{m!} N(E'-(m+1) E_{s}|\sigma)\\
    &=\sum_{m=1}^{\infty} \frac{\lambda_{s}^{m-1} e^{-\lambda_{s}}}{(m-1)!} N(E'-m E_{s}|\sigma)
\end{split}    
\label{eq:hatf?_DM}
\end{equation}

Finally, we can calculate the contaminated net differential sensitivity at $E_{s}$ 
\begin{equation}
\begin{split}
   &\Delta f(E'|S+E_{s})=- e^{-\lambda_{s}} N(E'|\sigma) \\
   &+ \sum_{m=1}^{\infty} \left( {1- \frac{\lambda_{s}}{m}} \right) \frac{\lambda_{s}^{m-1} e^{-\lambda_{s}}}{(m-1)!} N(E'-m E_{s}|\sigma)
\label{eq:df?_DM}
\end{split}
\end{equation}

\begin{figure*}
    \centering
    \includegraphics[width=0.75\textwidth]{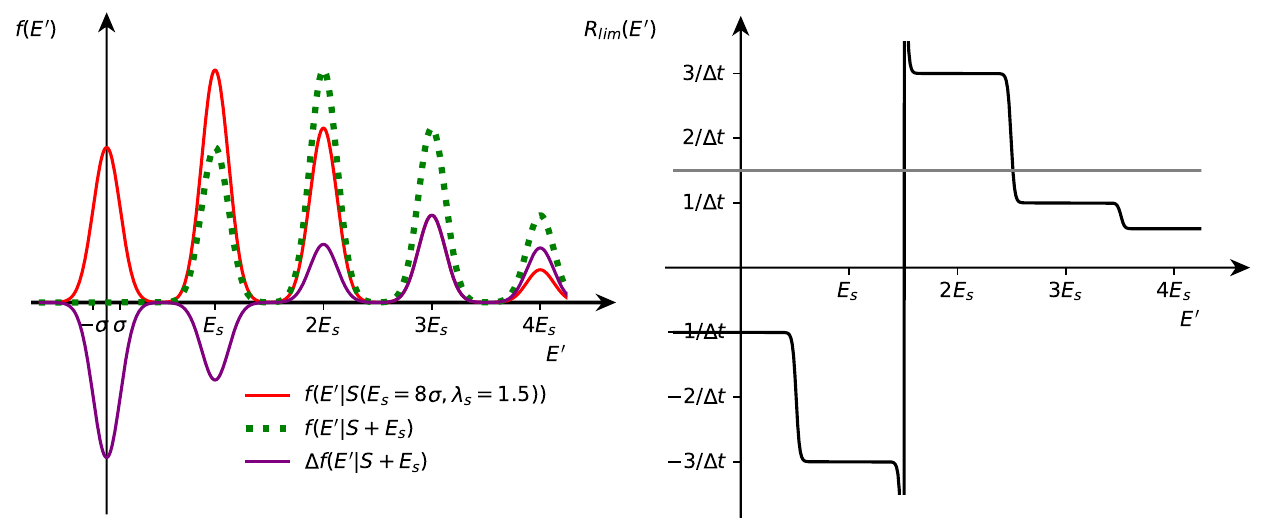} 
    \caption{Left: probability distributions for the case where there is DM contamination of the background data (see text). Here, the unknown true dark matter rate $\lambda_s = 1.5$, and the net response is negative at the single $E_s$ peak. Right: limit set by eq.~\ref{eq:R_lim_largelambda_largeEs}. The gray line indicates the true DM background rate. The limit is conservative only when $1.5E_s<E'<2.5E_s$.}
\label{fig:f_DMcontamination}
\end{figure*}
As shown in Fig.~\ref{fig:f_DMcontamination},  for $m < \lambda_{s}$, $\Delta f(E'|S+E_{s})$ is negative, while for $m> \lambda_{s}$ it's positive. This is a true effect that's qualitatively similar to the fact that the probability of measuring a bin with no DM interactions decreases as we increase the signal rate (Sec. \ref{sec:Discritized_SimpleGaussianNoise}). Integrating both $f(E'|S+E_{s})$ and $\Delta f(E'|S)$ over the individual, discrete peaks we find that 
\begin{equation}
F(m|S)= \frac{\lambda_{s}^{m} e^{-\lambda_{s}}}{m!}
\end{equation}
and 
\begin{equation}
\Delta F(m|S+E_{s})= \left( {1- \frac{\lambda_{s}}{m}} \right) \frac{\lambda_{s}^{m-1} e^{-\lambda_{s}}}{(m-1)!}
\end{equation}
for $m\geq1$ and generate a limit from each peak using Eq. \ref{eq:dRdE'_conservative}: 
\begin{equation}
\begin{split}
\widehat R_\mathrm{lim}(m|E_{s}) &\xrightarrow{T_\mathrm{Total}\to\infty}  \frac{1}{\Delta t} \frac{F(m|S)}{\Delta F(m|S+E_{s})}\\
&= \frac{R_{s}}{m-\lambda_{s}}
\end{split}
\label{eq:R_lim_largelambda_largeEs}
\end{equation}
and thus only the peak where $0\leq m-\lambda_{s}\leq 1$ is conservative and overestimates $R_{s}$. \textbf{All peaks with $ m-\lambda_{s}\geq 1$ will underestimate $R_{s}$ and be non-conservative. } 

This example highlights that a small signal first order taylor expansion of $f(E'|S)$ is obviously not accurate when the signal rate is large enough to produce significant pileup. A slightly deeper statement, is that in the small signal limit,
$f(E'|S)$ is a good approximation to $f(E'|0)$ (Eq. \ref{eq:f(E'|0)_FromSearch)})
and $f(E'|S+E_{s})$ is a good approximation to $f(E'|E_{s})$ (Eq. \ref{eq:f(E'|nsEs)_FromSearch}). At high interaction rates, this simply isn't true. 

\subsubsection{ \texorpdfstring{$\lambda_{s}\gg1$}{lambda s >> 1} with Gaussian Noise where \texorpdfstring{$\sigma \gg E_{s}$}{sigma o << E s}}
\label{sec:ex_largelambda_smallEs}
The previous example is quite specific to idealized integrating detectors with a Dirac-delta signal because pileup in  waveform detectors will not produce a quantized spectrum. However, it emphasizes that even in the most ideal scenarios, signal pileup can lead to non-conservative limits.

By contrast, the opposite limit where $\lambda_{s}\gg1$ and  $\sigma\gg E_{s}$ remains qualitatively similar for waveform detectors and thus significantly guides our intuition. In this limit, the Poissonian signal distribution can be well approximated as an off-centered Gaussian distribution,
 $N(E'-\lambda E_{s}| \sigma_{s}=\sqrt{\lambda_{s}}E_{s})$.
 Then when convoluted with Gaussian noise spectrum the various $m$ signal event coincidences gets smeared and strongly overlap leading to a smooth, non-quantized distribution:
\begin{equation}
     f(E'|S) = N(E'-\lambda_{s} E_{s}|\sigma=\sqrt{\sigma_{o}^2+\lambda_{s} E_{s}^2})
 \label{eq:GaussianSmallEs}    
 \end{equation}
where $\sigma_o$ is the original detector resolution without any events.
 Since $E_{s} \ll \sigma$, $\Delta f(E'|S+E_{s}) \sim -\frac{\partial f(E'|S)}{\partial E'} E_{s}$ 
and therefore 
\begin{equation}
\widehat R_\mathrm{lim} = \frac{-1}{\Delta t E_{s}} \frac{f(E'|S)}{\frac{\partial f(E'|E_{s})}{\partial E'}} = \frac{-1}{\Delta t E_{s}} \frac{1}{\frac{\partial \ln{f(E'|S)}}{\partial E'}}
\end{equation}
Plugging in Eq.\ref{eq:GaussianSmallEs}, we find that 
\begin{equation}
\begin{split}
&\widehat R_\mathrm{lim}(E'| E_{s})) \xrightarrow{T_\mathrm{Total}\to\infty} \\&\frac{\sigma_{o}^{2}}{\Delta t  E_{s} (E'- \lambda_{s}E_{s})} 
+ \frac{R_{s}}{\frac{E'}{E_{s}}-\lambda_{s}}
\end{split}
\end{equation}
and we end up again with the unfortunate conclusion that if the underlying Gaussian noise is dominated by Poisson noise fluctuations of the signal (i.e., $\sigma_{o}^2\ll \lambda_{s}E_{s}^2$), then $R_\mathrm{lim}(E'|E_{s})$ will underestimate $R_{s}$ by the factor $\frac{E'}{E_{s}}-\lambda_{s}$. 

This explicit test case really highlights the most important scenario for which using the net linear differential signal response (Eq.~\ref{eq:dRdE'_conservative_linear2}) to set an upper limit can be non-conservative; high signal pileup ($\lambda_{s}\gg1$) is largely indistinguishable from noise fluctuations and thus there is no way to determine if the signal is being boosted by true noise/background instead of other pileup signal interactions when $E_{s}\ll \sigma$. To both determine the relevant conservative range precisely and set conservative limits on signals where $E_{s}\ll \sigma$, we suggest using methods that set signal limits based upon the noise distribution itself \cite{das2024dark}.

\section{Generalization to a spectrum of energy deposition}
\label{sec:digitized device generic}
In the previous two sections, we have worked with the case where the putative signal was expected to deposit a single true energy in the detector, e.g., to take the example of DM, a simple Bosonic DM interaction. This simplified the calculation and allowed us to gain intuition. In this section, we extend the discussion to the general case of DM interactions with a continuous true energy deposition spectrum. 

\subsection{The case for a generic dark matter interaction}
Before we consider a generic DM interaction, let us first consider the signal model of multiple discrete true energies $E_{si}$, with rate $R_{si}$ respectively. We would have to take into account not only the pile-ups between the same energies as in Eq.~\ref{eq:f(E'|S)_full3}, but also pile-ups between different energy $E_{si}$ and $E_{sj}$, and triple pile-ups between $E_{si},~E_{sj}$ and $E_{sk}$, and so on. The formulae get rapidly out off hand but we can notice that any of these pile-up terms are at least of order two either in $R_{si}\Delta t\equiv\lambda_{si}$ or products of several rates $\lambda_{si},~\lambda_{sj},...$, which are also at least of order two in rates. Therefore, if we limit ourselves to the first-order terms in rates, which is the case of negligible pile-ups, Eq.~\ref{eq:f(E'|S)'_linear} becomes
\begin{equation}
\begin{split}
   &f(E'|S(\{E_{si},R_{si}\})) \\
   &=f(E'|0) +\sum_{i}R_{si}\Delta t(f(E'|E_{si})-f(E'|0))
\end{split}
\label{eq:f(E'|S)'_linear-mul}
\end{equation}

Now, in the no pile-up approximation, for interaction processes that produce a continuum of energy depositions/signal magnitudes as found in DM scattering, one can easily generalize Eq.~\ref{eq:f(E'|S)'_linear} and Eq.~\ref{eq:f(E'|S)'_linear-mul} by simply integrating over the true energy, $E_{s}$:
\begin{widetext}
\begin{equation}
f( E'|S(M_{s},\sigma_{s}) ) = f (E'|0) +  \Delta t \int dE_{s} \frac{dR}{dE_{s}}\left(E_{s}|S(M_{s}, \sigma_{s}) \right) \Bigl( f(E' |E_{s})- f(E' |0)\Bigr)+... 
\end{equation}
where $\frac{dR}{dE_{s}}(E_s|S)$ is the differential rate of DM with mass $M_s$ at the true but unknown cross-section $\sigma_s$. The common practice is to calculate DM spectrum shape $\frac{dR}{dE_{s}}(E_{s}|S)/R(S)$ at an arbitrary reference cross-section $\sigma_0$, and move the linear dependence on $\sigma_s$ to the total signal rate $R_{s}(S(M_{s},\sigma_{s}))$:
\begin{equation}
\begin{split}
&f( E'|S(M_{s},\sigma_{s}) ) \\
&= f (E'|0) +  R_{s}(S(M_{s},\sigma_{s})) \Delta t \int dE_{s} \frac{\frac{dR}{dE_{s}}\left(E_{s}|S(M_{s}, \sigma_{0}) \right)}{R(S(M_{s},\sigma_{0}))} \Bigl( f(E' |E_{s})- f(E' |0)\Bigr)+... \\
&= f (E'|0 ) +  R_{s} \Delta t \Delta f(E' |s(M_{s}))+ ...
\end{split}
\label{eq:dRdE'int_0}
\end{equation}
In essence, in Eq.~\ref{eq:dRdE'int_0} we are simply weighting the various $\Delta f(E'|E_{s})$ by the shape of the true recoil spectrum to calculate $\Delta f(E'|s(M_{s}))$, the net differential sensitivity for a single elastic scatter with mass $M_{s}$. $s(M_s)$ is the generalization of salting with a true energy deposition $E_s$, where one salts with true energy deposition distribution whose shape is parameterized by the DM mass, $M_{s}$. As with Eq.~\ref{eq:dRdE'_linear}, this equation is only valid in the limit of  minimal signal pileup, $(\lambda_{s}=R_{s}\Delta t \ll 1$). 

Just as for the case of Bosonic DM, one can also estimate the true net differential signal sensitivity by measuring the potentially signal-contaminated distributions, provided the true $\lambda_{s}\ll1$:
\begin{equation}
\begin{split}
&f( E'|S(M_{s},\sigma_{s})) = f (E'|0)\\
&+ R_{s}(S(M_{s},\sigma_{s})) \Delta t \int dE_{s} \frac{\frac{dR}{dE_{s}}(E_{s}|S(M_{s},\sigma_{0}))}{R(S(M_{s},\sigma_{0}))} \Bigl( f(E' |S(M_{s},\sigma_{s})\!+\!E_{s})- f(E' |S(M_{s},\sigma_{s}))\Bigr)+... \\
&= f (E'|0 )\! + \! R_{s} \Delta t \Delta f(E' |S(M_{s},\sigma_{s})+s)+ ...
\end{split}
\label{eq:dRdE'int}
\end{equation}
\end{widetext}
where the dropped higher order terms scale as $\lambda_{s}^2$ or greater. This equation can be understood as the generalization of Eq.~\ref{eq:dRdE'_conservative_linear1}. And the DM signal upper limit can be estimated in the same form as Eq.~\ref{eq:dRdE'_conservative}.

\subsection{Simulated Light Mass Dark Matter Search}
\label{sec:digitized device sim DM search}
In order to give a semi-realistic example of what a DM search could look like, we explore a scenario with \SI{50}{MeV} DM from a standard halo velocity distribution that interacts via scalar nuclear elastic scattering with a \SI{5.6e-33}{\square\centi\meter} cross section using a \SI{10}{\gram} Si ideal integrating detector that has an intrinsic Gaussian noise of \SI{1}{\electronvolt}$_\mathrm{rms}$. The same event and backgrounds are observed with two different integration times, one at \SI{100}{\micro\second} per frame and one $\times100$ slower, at \SI{10}{\milli\second} per frame. The resulting difference in $\lambda_s$ creates drastically different measured experimental differential rates and transitions us from having noise-boosted limits that are guaranteed to be conservative (Sec.~\ref{sec:DMlim_Conservative}) to the high pileup regime where it's possible that the derived limits could be non-conservative (Sec. \ref{sec:DMlim_pileup}). 

\begin{figure*}
    \centering
    \includegraphics[width=0.70\linewidth,trim={0 0.2cm 0 0},clip]{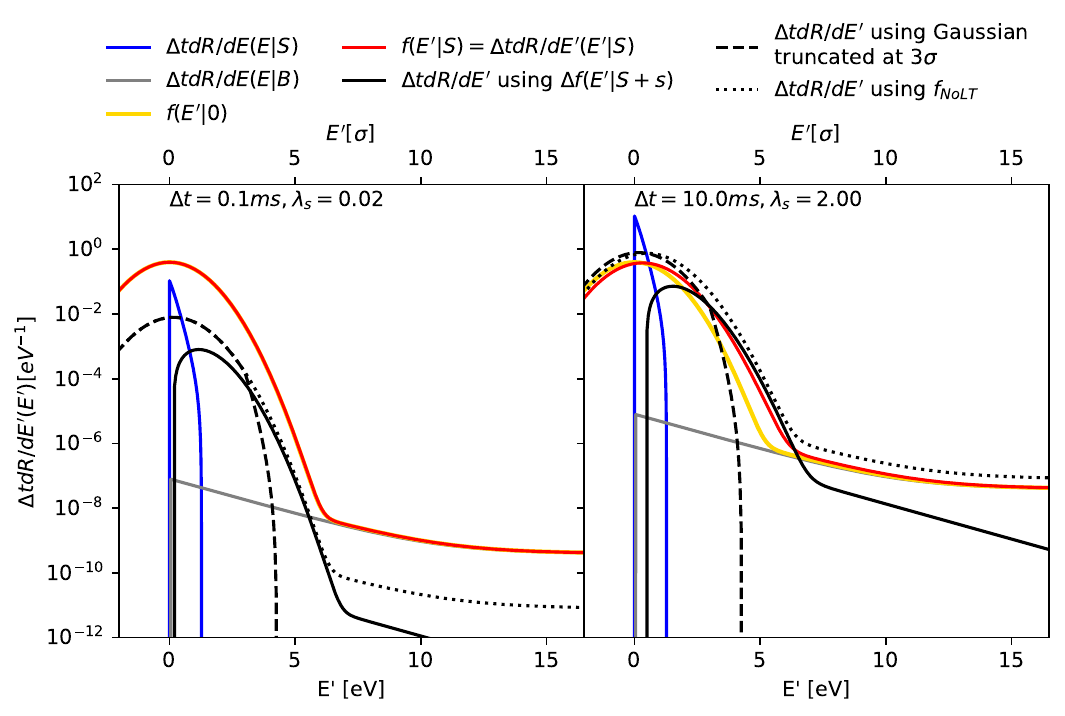}
    \caption{A realistic detector background spectrum without dark matter, yellow, $f(E'|0)$, and with the presence of unknown DM signals, red, $f(E'|S)$. 
    The background spectrum is shown in gray.
    The unknown signal, blue, is from \SI{50}{\mega\electronvolt} nuclear recoil DM with $\sigma_{n}=\SI{5.6e-33}{\square\centi\meter}$ scattering in a \SI{10}{\gram} silicon detector. The resulting expected signal per integration time, $\lambda_s$, is 0.02 (2) for the fast (slow) integration time. The detected DM spectrum $dR/dE'$, black, is estimated using different methods. The solid black line uses the correct net differential rate change proposed in this work, Eq.~\ref{eq:dRdE'int}, the dotted black line uses the smearing without signal live-time correction, Eq.~\ref{eq:dRbaddE'}, and the dashed line is smeared assuming Gaussian baseline fluctuation and truncated at $3\sigma$, Eq.~\ref{eq:dRdE'_3sigma}. }
    \label{fig:DMsim_dRdE}
\end{figure*}

The true energy differential signal rate, $\frac{dR}{dE}(E|S(M_{s},\sigma_{n}))$, where $S$ represents the unknown signals from DM of $M_{s}=\SI{50}{\mega\electronvolt}$ and $\sigma_n=\SI{5.6e-33}{\square\centi\meter}$, is shown in Fig. \ref{fig:DMsim_dRdE} scaled by the integration time $\Delta t$. We explicitly note that unlike $\frac{dR}{dE'}(E') \Delta t = f(E')$ that has an integral of 1,  the integral of $\frac{dR}{dE}(E|S) \Delta t$ is $\lambda_{s}$ and therefore is not a probability distribution function. The signal cross section was purposefully chosen such that $\lambda_s=0.02$ for the detector with the faster integration time (left) and $2.0$ for the detector with the slower integration time (right). 
The true energy background differential rate, $\frac{dR}{dE}(E|B)$, was chosen to be qualitatively similar in shape to the low energy excess \cite{10.21468/SciPostPhysProc.9.001} commonly seen in the current generation light mass DM calorimeters but scaled to not completely dominate the signal. 

Next, we construct the experimentally measured probability with the true signal and background events hitting the detector,  $f(E'|S)$ (red). For the fast integrating detector, $\lambda_{s}$ is small, and the first order Taylor expansion formulation, Eq.~\ref{eq:dRdE'int}, can be used. For the slowly integrating detector, a continuum true energy deposition generalized form of Eq.~\ref{eq:dRdE'_full3} is needed since there is significant signal pileup. To highlight the changes in the spectrum from the signal interaction, we also plot the unmeasurable background-only distribution, $f(E'|0)$ (yellow).  

 The high energy tail of $f(E'|S)$ and $f(E'|0)$ for both detectors is simply a convolution of the noise point spread function ($\sigma =\SI{1}{\electronvolt}$) with the true $\frac{dR}{dE}(E|B)$. Since the background event distribution is relatively slowly varying compared to the noise energy scale, $f(E'|S)$ and $f(E'|0) \sim \frac{dR}{dE}(E|B) \Delta t$ in the high energy range. 

 At low measured energies, there are substantial differences in the performance of the two detectors. For the quick detector,  differences between $f(E'|S)$ and $f(E'|0)$ do exist but are difficult to see since only $2\%$ of the bins have a signal interaction, and even then, a single signal interaction event has an average energy deposition $\sim \sigma$. For the slow detector, however, there is a significant noise broadening for $f(E'|S)$ due to DM signal shot noise as derived and discussed in Sec.~\ref{sec:ex_largelambda_smallEs} compared to $f(E'|0)$
 
\begin{figure*}
    \centering
    \includegraphics[width=0.70\linewidth,trim={0 0.2cm 0 0},clip]{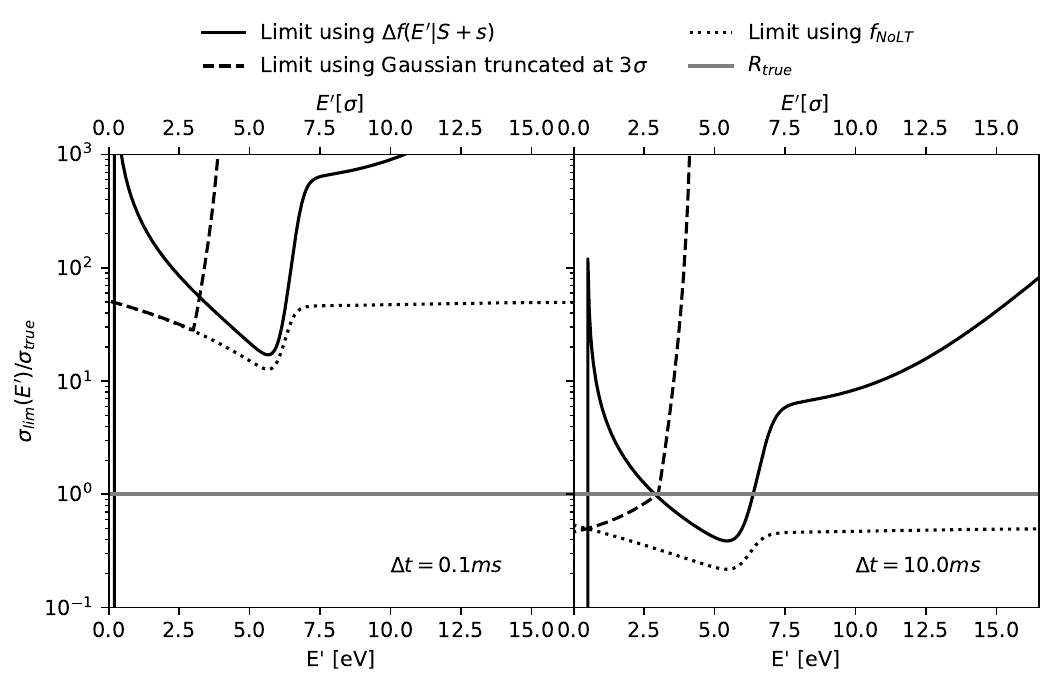}
    \caption{Limit of dark matter rate as a function of the measured energy $E'$ with respect to the true DM background rate, assuming infinite exposure. Line styles are the same as in fig.~\ref{fig:DMsim_dRdE}. Limits in the left figure are conservative. In the right figure, long frame time results in $\lambda_s>1$, and the limits are underestimated at certain $E'$.}
    \label{fig:DMsim_limit}
\end{figure*}

Positive portions of the linear net differential sensitivity for DM nuclear scattering, $\Delta f(E'|S+s)$ (Eq.~\ref{eq:dRdE'int}) with $M_{s}=\SI{50}{\mega\electronvolt}$ are shown in solid black, and the resulting sensitivity limits shown in Fig.~\ref{fig:DMsim_limit} are produced with the generalization of Eq.~\ref{eq:dRdE'_conservative}:
\begin{equation}
\widehat{R}_\mathrm{lim}\equiv \frac{1}{\Delta t}\frac{\hat{f}(E'|S)}{\widehat{\Delta f}(E'|S+s)} 
\label{eq:Rlim_conservative_generalized}
\end{equation}

For the fast detector with minimal pileup, the resulting limits (solid black) are found to be conservative at all measured energies as proven in Sec. \ref{sec:DMlim_Conservative}. By contrast, for the slow detector with $\lambda_{s} = 2$, the resulting limits are actually smaller than the true signal rate in the range from \SI{2.5}{\electronvolt} to \SI{7}{\electronvolt}, illustrating the danger and non-conservativeness of using limit estimators that assume minimal pileup.  

\subsection{Techniques to guarantee the conservativeness for all background and signal scenarios}
We've now given explicit examples where the calculated limit will be lower than the true rate for various Dirac-delta (Sec.~\ref{sec:DMlim_pileup}) and continuous spectra signals (Sec.~\ref{sec:digitized device sim DM search}) in scenarios when the measured differential rate is dominated by true signal pileup. The most elegant way to guarantee conservativeness of this limit estimator is to first exclude the problematic high signal pileup space with a less sensitive but manifestly conservative estimator. For example, \cite{das2024dark}, and further improvements on this concept \cite{NoiseLimit-work-in-progress}, set absolutely conservative limits on the noise distribution itself. 

 Alternatively, we can constrain the measured energy estimator domain from which we can build our limit estimator. As shown in Fig.~\ref{fig:DMsim_limit}, problematic non-conservative performance of the estimator ($\SI{3}{\electronvolt}\sim\SI{7}{\electronvolt}$) occurs in the energy range where the measured differential rate is dominated by Gaussian noise from true signal pileup. If one constrains the domain of limit generation only to non-gaussian regions only, the resulting limit should be conservative.

 We also emphasize that the exclusion regions produced by linear net differential sensitivity analyses should include the pileup rate as an upper contour edge.

\subsection{Conservativeness of Previous Experiments}
\label{sec:PrevExpDiscussion}
As discussed earlier,  Eq.~\ref{eq:dRdE'int} was not used to estimate $\frac{dR}{dE'}$ in recent noise boosting analyses. Instead the CPDv1 DM search \cite{CPDv1_PRL21_LDMSearch}, the EDELWEISS search \cite{EDELWEISS_PRD19_LDMSearch}, and the CRESST searches \cite{CRESST3_PRD19_FirstDMLimits, CRESST_PRD24_SOS_DMLimits} did not account for the decrease in non-signal live-time and used
\begin{widetext}
\begin{equation}
\begin{split}
f_\mathrm{NoLT}(E'|S(M_{s},\sigma_{s})) & = f(E'|0) +  R_{s}(S(M_{s},\sigma_{s})) \Delta t \int dE_{s} \frac{\frac{dR}{dE_{s}}(E_{s}|S(M_{s},\sigma_{0}))}{R(S(M_{s},\sigma_{0}))} f(E' |S(M_{s},\sigma_{s})+E_{s})\\
                                        &= f(E'0) + R_{s} \Delta t f(E'|S+s)    
\end{split}
\label{eq:dRbaddE'}
\end{equation}
which will overestimate signal sensitivity when there is significant overlap between  $f(E'|S)$ and $f(E'|0)$. This can be explicitly seen in the example DM search shown in  Fig.~\ref{fig:DMsim_limit} where $f(E'|S+s)$ has been plotted (dotted black) and can be compared to the net differential sensitivity, $\Delta f(E'|S+s)$ (solid black). As discussed in Sec. \ref{sec:DMsearch_flattail}, for signals with small energy depositions (like in this example for a \SI{50}{\mega\electronvolt} DM nuclear recoil)  $\Delta f(E'|S+s) \propto \frac{d f(E'|S)}{dE'}$ and thus in the relatively flat non-gaussian outlier tail region   $\Delta f(E'|S+s)$ is correctly suppressed and $f(E'|S)$ substantially overestimates sensitivity.

Since this danger was partially understood, each of these searches  employed additional {\it ad hoc} constraints on the amount of noise boosting. To take one example, \cite{CPDv1_PRL21_LDMSearch} required that the boosted measured energy, $E'$, is no more than $3\sigma$ higher than the true energy $E$:
\begin{equation}
\begin{split}
    &f_{3\sigma\mathrm{lim}}(E'|S(M_{s},\sigma_{s})) \\
    &= f(E'|0)  + R_{s}(S(M_{s},\sigma_{s})) \Delta t \int dE_{s} \frac{\frac{dR}{dE_{s}}(E_{s}|S(M_{s},\sigma_{0}))}{R(S(M_{s},\sigma_{0}))}f(E'|E_{s}) H(E_s+3\sigma-E')
\end{split}
\label{eq:dRdE'_3sigma}
\end{equation}
\end{widetext}
where $H$ is the Heaviside step function. The resulting limits can be seen in Fig.~\ref{fig:DMsim_limit} (dotted black) for the case of \SI{50}{\mega\electronvolt} DM nuclear elastic scattering. For this specific scenario, the 3$\sigma$ constraint in combination with the 4.5$\sigma$ trigger threshold means, by construction, there is absolutely no experimental sensitivity to all signals with an energy deposition below 1.5$\sigma$. This combination of trigger thresholds and cuts is designed to guarantee that the experiment didn't overestimate its sensitivity to very light mass DM, but it is overly conservative for the lightest DM models, and the transition to limits based on the net differential sensitivity will produce stronger but still conservative DM limits.

We can also attempt to assess the conservativeness of these historical searches for higher mass DM where $E_{s}> 1.5 \sigma$. For $E' > 4.5 \sigma$, 
$f(E'|0)$ is relatively small. Therefore,
$\Delta f(E'|E_{s}) = f(E'|E_{s})-f(E'|S) \sim f(E'|E_{s})$ as long as $E_{s} > 1.5 \sigma$.  Consequently not accounting for the decreased no signal live-time, shouldn't be problematic for high mass DM nuclear recoil scatters with an average $\langle E \rangle > 1.5  \sigma$. The conservativeness of the intermediate cross over range is less clear. Potentially there is a small mass range in which only a very small fraction of the hypothetical DM nuclear scatters are above 1.5$\sigma$. In this small mass region, the approximations made could theoretically overestimate DM sensitivity. However, since it is known that the signal in this range is dominated by a time varying non-ionizing background source that isn't DM \cite{CRESST3_PRD19_FirstDMLimits}, these limits are almost certainly conservative in this intermediate range as well. For \cite{CRESST_PRD24_SOS_DMLimits} and \cite{EDELWEISS_PRD19_LDMSearch} the actual {\it post hoc} cuts and trigger thresholds are quantitatively different and even a bit more conservative, but qualitatively their performance is identical.

In the future, all near threshold signal searches should really build exclusion contours from the net differential sensitivity to maximize their small signal sensitivity while still guaranteeing conservativeness for all signal sizes as long as there is minimal pileup.

 \section{Understanding waveform Detectors}
 \label{sec:waveform_detector}

 So far, we have correctly calculated $\frac{dR}{dE'}$ as a function of $R_{s}$ for only an idealized integrating time-binned detector like a CCD. In this section, we will qualitatively discuss the steps in turning a continuous time stream from an waveform detector into a set of events above a trigger threshold, and show that the simplified model we developed is still qualitatively accurate. In particular, the fact that an increase in signal rate both increases the rate of noise+signal coincidence and \textbf{decreases the noise-only event rate} is simply a consequence of the fact that the total data acquisition time is fixed regardless of detector subtleties.

A modern readout that maximizes signal to noise will take an analog continuous stream of data, demix the stream if necessary, Nyquist filter, and then digitize the stream at a high enough rate that nearly the entire information content of the signal remains with high fidelity (a good rule of thumb is that the Nyquist frequency is at least $\times10$ larger than the fundamental dynamical poles of the signal referenced to detector output for small signals and $\sim\times10$ than the largest dynamical pole for optimum large signal fidelity). An acausal optimum filter or matched filter is then applied to this digitized stream to calculate a signal amplitude estimator for an event that occurs at $t$, $E'(t)$.

A trigger algorithm will then go through this time stream and find regions where $E' > E'_{t}$, the trigger threshold energy. Each above-threshold region is then converted to a discrete number of energy depositions occurring at precise times. The simplest possible trigger algorithm, for example, is to associate a single signal event with each above threshold region that has a signal amplitude estimator, $E'$, and an event time estimator, $t'$ that occurs at the relative maximum of $E'(t)$ in the above threshold region. 

After acquiring the triggered regions, various additional selections are made to cull events that aren't consistent with high-quality signal events (pileups, saturated events, events occurring in periods of poor detector performance, events with pulse shapes inconsistent with true signal events, etc.).

\subsection{Optimum Filter Signal Amplitude Estimator} 
\label{sec:waveform_OF}
The simplest optimum or matched filter \cite{zadeh1952optimum} generates a signal amplitude estimator for an event occurring at $t$ with the highest signal-to-noise ratio on an experimental trace, $Y(\omega)$, provided that:
\begin{itemize}
    \item the noise is stationary, Gaussian distributed and characterized by a measured noise variance, $\sigma^2(\omega)$
    \item the signal shape is constant for all events regardless of size and is well modeled by a template, $T(\omega)$.
    \item at most, there exists only a single event within the trace. 
\end{itemize}

A derivation for the optimal estimator follows from maximizing the probability likelihood or equivalently minimizing the $\chi^{2}$ of the residual
\begin{widetext}
\begin{equation}
\chi^{2} = \sum_{\omega} \left(Y^{*}(\omega)- E' T^{*}(\omega) e^{j\omega t}\right)\sigma(\omega)^{-2}(Y(\omega)- E' T(\omega) e^{-j\omega t})
\label{eq:Chi2}
\end{equation}
with respect to $E'$ if the event time is known (as in the case during detector calibration with an LED pulse) and the resulting best-fit amplitude is 
\begin{equation}
	E'(t)= \frac{ \sum_{\omega} T^{\dagger}(\omega) \sigma(\omega)^{-2} Y(\omega) e^{j \omega t}}{\sum_{\omega}T^{\dagger}(\omega) \sigma(\omega)^{-2} T(\omega) }
 \label{eq:OF}
\end{equation}

\begin{figure*}
    \centering
    \includegraphics[width=0.80\textwidth]{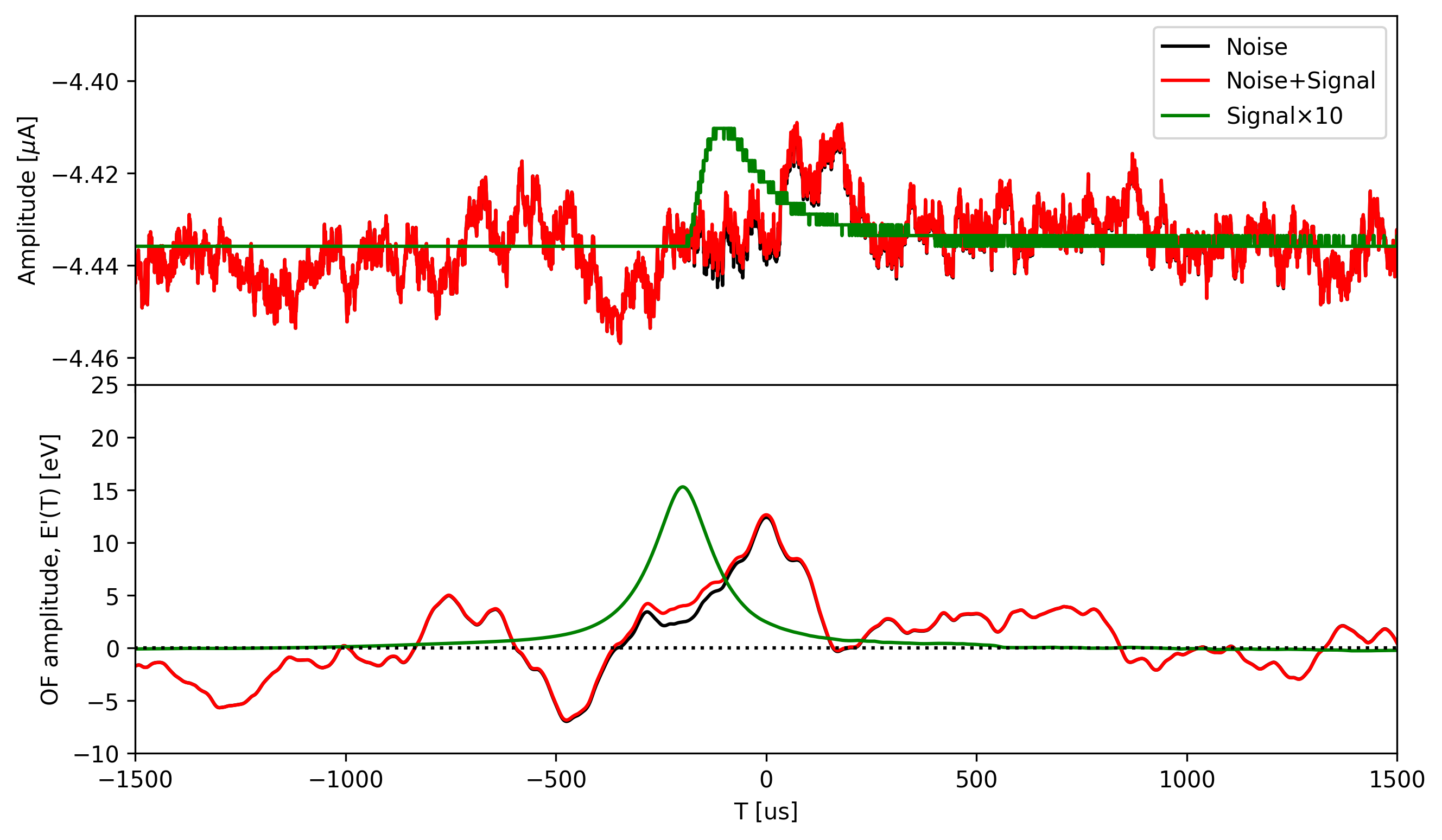} 
    \caption{ Optimal filter example: A small true signal event (green, scaled by 10) is in coincidence with various noise fluctuations (black), producing the summed coincidence trace (Red). The top panel shows traces \textbf{before} optimal filtering, while the bottom panel displays the traces \textbf{after} optimal filtering. In the bottom panel, the black trace is $E'_{n}(t)$ and the green trace is $E_{s} \rho(|t-t_{s}|)$ (scaled by x10). Finally, the red curve is the optimum filter output of the salted trace, which shows the behavior that we derived in eq. \ref{eq:OFsalt}.}
\label{fig:OFtrace}
\end{figure*}
To understand how a hypothetical true signal impacts $E'(t)$, we can replace $Y$ with the sum of a noise trace, $Y_{n}$, and a true signal event that occurs at time $t_{s}$ with amplitude $E_{s}$ and pulse shape $T(\omega)$ as shown graphically in  Fig. \ref{fig:OFtrace}. After Optimum filtering, we find
\begin{equation}
\begin{split}
	E'(t) 
	        =& \frac{ \sum_{\omega} T^{\dagger}(\omega) \sigma(\omega)^{-2} (Y_{n}+E_{s}T(\omega)e^{-j\omega t_{s}})e^{j \omega t}}{\sum_{\omega}T^{\dagger}(\omega) \sigma(\omega)^{-2} T(\omega)}\\
	        =&  \frac{ \sum_{\omega} T^{\dagger}(\omega) \sigma(\omega)^{-2} Y_{n} e^{j \omega t}}{\sum_{\omega}T^{\dagger}(\omega) \sigma(\omega)^{-2} T(\omega)}+E_{s}  \frac{ \sum_{\omega} T^{\dagger}(\omega) \sigma(\omega)^{-2} T(\omega) e^{j \omega (t-t_{s})}}{\sum_{\omega}T^{\dagger}(\omega) \sigma(\omega)^{-2} T(\omega)}\\
	        =& E'_{n}(t)+E_{s} \rho(|t-t_{s}|)
\end{split}	
\label{eq:OFsalt}
\end{equation} 
\end{widetext}
where $E'_{n}(t)$ is the noise only trace energy estimator and 
 $\rho(|t-t_{s}|)$ is the normalized weighting function which encapsulates how a true signal at $t_{s}$ affects $E'$ at t. $\rho$'s behavior largely depends upon the precise noise and template used. However, there are two completely general properties. First, at $t=t_{s}$,  $\rho(0)=1$. Secondly, $|\rho(|t-t_{s}|)|<=1$ for all t. The former means that when estimating  $E'(t=t_{s})$ the filter behaves like the perfect simplistic discrete integrator: $E'(t_{s})=E_{n}'(t_{s})+E_{s}$. For all other t ($t \neq t_s$), the effect of the signal is attenuated. 

If the noise and template have no peaks in the frequency space, then additionally $\rho(|t-t_{s}|)$ decreases monotonically with $|t-t_{s}|$ and $\rho(|t-t_{s}|)$ effectively defines a qualitative digitization scale $\Delta t_\mathrm{OF}$ below which the noise and signal qualitatively sum, and above which the change is qualitatively negligible. This behavior is qualitatively similar to the boxcar function response of the idealized integrating detector (CCD) we used in Sec.~\ref{sec:digitized device} where an event at time $t_s$ is sensed without attenuation by the time bin in which it is located but doesn't affect measurements of other time bins:
\begin{equation}
\rho_\mathrm{CCD}(t,t_s) =
\left\{
	\begin{array}{ll}
            0  & \mbox{for } t_{s}<t_{j} \\
		1  & \mbox{for } t_{j}< t_{s}<t_{j+1} \\
            0  & \mbox{for } t_{j+1}<t_{s}
	\end{array}
\right.
\end{equation}

Somewhat surprisingly, $\rho(t)$ is also the Pearson correlation coefficient (the normalized autocorrelation function) for the optimum filtered traces, and thus $\rho$ is a measurement of the correlation time scale on the noise, $E'_{n}(t)$. If two random times, $t_{1}$ and $t_{2}$, are close temporally ($|t_{1}-t_{2}| < \Delta t_\mathrm{OF}$), then  $E'_{n}(t_{1})$ and $E'_{n}(t_{2})$ will be strongly correlated. By contrast, for $|t_{1}-t_{2}| > \Delta t_\mathrm{OF}$,  $E'_{n}(t_{1})$ and $E'_{n}(t_{2})$ will become increasingly weakly correlated as the time difference increases. This is again qualitatively like the idealized CCD where the noise in the $j$ and $j+1$ bins are uncorrelated. \textbf{In summary, an waveform detector acts qualitatively similar to the idealized integrating detector with an effective $\Delta t_\mathrm{OF}$ and thus the intuition that we developed with idealized integrating detectors is still valid.}

There are two more non-intuitive deviations from perfect qualitative correspondence to the idealized integrating detector that we regularly encounter when implementing optimum filters.  First, as foreshadowed above, sharp peaks at a given frequency in the noise or signal template, for example, a large \SI{60}{\hertz} peak due to EMI pickup, will produce a $\rho(|t-t_{s}|)$ that is non-monotonic and has multiple relative minima or ``echoes". 
Secondly, we regularly find that our DC noise is significantly larger than our AC noise terms. In fact, we usually artificially set the DC noise to $\infty$ to minimize propagation of very long time scale changes in the detector equilibrium into our energy estimator. This lack of a DC term in the optimum filter means that the time average of the weighting function, $\langle\rho(|t-t_{s}|)\rangle_{t}=0$. Consequently, for $|t_1-t_2| \gg\Delta t_{OF}$, $\langle \rho (t_1-t_2)\rangle <0$; counter-intuitively, the addition of a positive true signal far from a noise peak will actually slightly decrease $E'$. 

\subsection{Choosing Trace Length for the Optimum Filter}
To our knowledge, there isn't a clear optimization strategy for choosing the trace length, $T_\mathrm{trace}$, used in generating the Optimum filter. On the one hand, larger $T_\mathrm{trace}$ gives one greater frequency resolution and thus the optimum filter has improved ability to optimally weight different frequencies for improved sensitivity which is extremely useful in removing  sharp environmental noise peaks like \SI{60}{Hz}. On the other hand, larger trace lengths mean a higher probability of pileup of both signal and background events which breaks one of the assumptions required for optimality of the estimator. Thus, the total background rate and the needed calibration rate set a soft upper bound on $T_\mathrm{trace}$ since having $1/T_\mathrm{trace} \gtrapprox$ to the signal+background+calibration rate will significantly increase the complexity and/or the live time loss. Consequently, $T_\mathrm{trace}$, is commonly chosen to be the shortest possible length that doesn't significantly degrade the resolution (by say $\sim 5 \%$).

\subsection{Estimating Event Time In Trigger Algorithm}
\label{sec:triggermodel}
Finally and most importantly, the distillation of $E'(t)$ into an event time estimate has no direct waveform in an integrating detector and thus it unfortunately adds significant additional complexity that must be at least qualitatively understood. If neither the event time nor amplitude is known for the signal, the $\chi^{2}$ is minimized with respect to both $E'$ and $t$. Since $\frac{\partial \chi^2}{\partial t}=0$ occurs at the relative extrema of $E'(t)$, the standard event time estimation is at the maximum $E'(t)$ in some fixed time window $T_\mathrm{merge}$ after the trigger goes above the trigger threshold. 

We want to specifically emphasize that selecting the maximum from a random distribution is an explicitly non-linear algorithm and thus its effects must be at minimum qualitatively modeled and understood. Since only one event is recorded within the $T_\mathrm{merge}$ window, the larger the $T_\mathrm{merge}$, the higher the probability of not triggering on an event either due to pileup with a larger event or a larger noise fluctuation within the merge window. Thus, in some ways $T_\mathrm{merge}$ acts as dead time. However, it is truly qualitatively different than a hardware-enforced trigger dead period; bigger events are more likely to be tagged and smaller events are more likely to be untagged. Minimizing these effects would suggest a smaller $T_\mathrm{merge}$ on the scale of $\Delta t_\mathrm{OF}$. On the other hand, sharp de-weighting of noise peaks in frequency space will tend to generate oscillation in the time domain, which for large enough pulses can be above the trigger threshold and thus generate echoes. If $T_\mathrm{merge}\sim \Delta t_\mathrm{OF} \ll T_\mathrm{trace}$, these echoes will be outside the merge window and will be triggered and tagged as real events; there is no perfect choice for simplistic trigger algorithms. 

As an intermediate step to full modeling of an waveform detector with an optimum filter trigger, let's follow \cite{EDELWEISS_PRD19_LDMSearch} and qualitatively model the effects of the trigger algorithm searching for the maximum $E'$ throughout a $T_\mathrm{merge}$. To do this, we will take our idealized integrating detector, but rather than store the measured signal in every bin, we will instead store only the maximum measured energy in every $N$ adjacent bins, which models a $T_\mathrm{merge}= N \Delta t_\mathrm{OF}$. For simplicity, we will also assume that the signal pileup probability,$\lambda_{s}=R_{s}\Delta t_\mathrm{OF}$, is small.

\begin{figure}
    \centering
    \includegraphics[width=\linewidth]{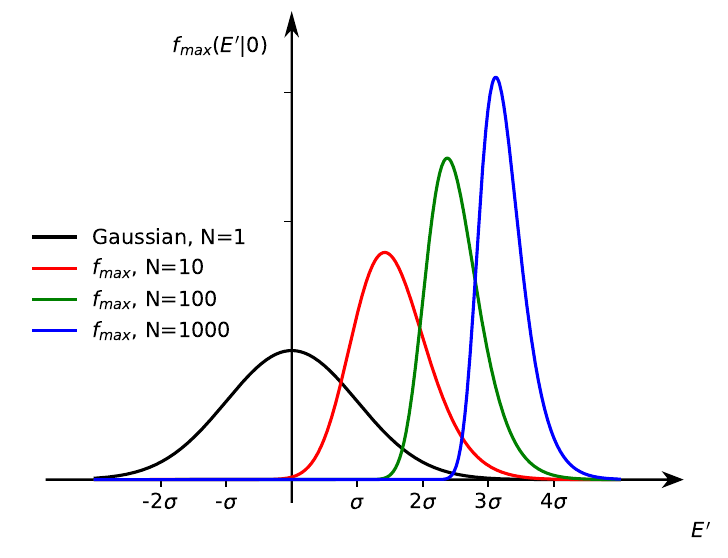}
    \caption{A normally distributed $f(E'|0)$ in black, and the resultant $f_\mathrm{max}(E'|0)$ for $N=10$ (red), 100 (green), and 1000 (blue)}
    \label{fig:fmax_E'}
\end{figure}

Redoing our original derivation for $\frac{dR}{dE'}$ in Eq. \ref{eq:dRdE'_full} with a $T_\mathrm{merge}$ window,  we find
\begin{widetext}
\begin{equation}
\begin{split}
\frac{dR}{dE'}(E'|S(E_{s},R_{s}))  \approx& \frac{1}{N \Delta t_\mathrm{OF}} \bigl(P(0| R_{s} N \Delta t_\mathrm{OF}) f_\mathrm{max}(E' | 0)+P(1| R_{s} N\Delta t_\mathrm{OF}) f_\mathrm{max}(E' | E_{s})\bigr) \\
					\approx&\frac{1}{N \Delta t_\mathrm{OF}} \bigl((1- R_{s} N \Delta t) f_\mathrm{max}(E' | 0)+  (R_{s} N \Delta t )f_\mathrm{max}(E' | E_{s})\bigr)\\
			        \approx&\frac{1}{N \Delta t_\mathrm{OF}}f_\mathrm{max}(E' | 0) +R_{s}\bigl( f_\mathrm{max}(E' | E_{s})-f_\mathrm{max}(E' | 0)\bigr) 
\end{split}
\label{eq:dRdE'_Tmerge}
\end{equation}
\end{widetext}
where $f_\mathrm{max}(E'|0)$ is the probability distribution of the maximum of the measured energies in $N$ bins, none of which have a signal interaction. Though it can be calculated directly, it's perhaps easiest to first calculate its cumulative distribution function, which is the probability that each and every bin has as energy $<E'$
\begin{equation}
F_\mathrm{max}(E'|0)= F(E'|0)^{N}
\label{eq:Fmax}
\end{equation}
where $F(E'|0)$ is the CDF of the individual bin energy distribution with no signal events and therefore
\begin{equation}
\begin{split}
f_\mathrm{max}(E'|0) &= \frac{dF_\mathrm{max}(E'|0)}{dE'}\\
              &= N f(E'|0)F(E'|0)^{N-1}
\end{split}
\label{eq:fmax}
\end{equation}
\bs{added this short justification}
In these equations (\ref{eq:Fmax} and \ref {eq:fmax}) we have made the simplifying assumptions that the $E'$ measurement in different $\Delta t_\mathrm{OF}$ intervals are independent and that the signal appears only in one of these intervals. This is approximately true with our choice of the time interval $\Delta t_\mathrm{OF}$ and thus these distributions are accurate enough to give qualitative insight into an optimal filter with $T_\mathrm{merge} > \Delta t_{\mathrm{OF}}$.

These approximations allowed us to use the standard methods to derive the  probability  distribution of maxima of independent random variables.

In Fig. \ref{fig:fmax_E'}, $f_\mathrm{max}(E'|0)$ is shown for a wide variety of merge window sizes for a normally distributed $f(E'|0)$. Following from Eq. \ref{eq:fmax}, we find that for $E'$ where $F(E'|0) < N^{-1/(N-1)}$, $f_\mathrm{max}(E'|0) < f(E'|0)$: the likelihood not triggering on the small noise/background event increases with increasing $N$ as intuitively expected and shown in the figure. Simply put, to have a low-energy noise/background event in the maximum distribution requires a low-energy noise/background event in every bin.  On the other hand, the probability of sampling a higher outlier tail event goes up substantially: in the limit as $F(E'|0) \rightarrow 1$, $f_\mathrm{max}(E'|0) = N f(E'|0)$. 

Likewise, $F_\mathrm{max}(E'|E_{s})$ can be derived by recognizing that the bin with the signal event and the $N-1$ bins without  the signal event must all have energies $< E'$,
\begin{equation}
F_\mathrm{max}(E'|E_{s}) = F(E'|E_{s}) F(E'|0)^{N-1}  
\label{eq:Fmax_Es}
\end{equation}
and therefore
\begin{equation}
\begin{split}
&f_\mathrm{max}(E'|E_{s}) = \frac{dF_\mathrm{max}(E'|E_{s})}{dE'}\\
              &= f(E'|E_{s}) F(E'|0)^{N-1}\\
              &+F(E'|E_{s}) (N-1)f(E'|0)F(E'|0)^{N-2}
\end{split}
\label{eq:fmax_Es}
\end{equation}
Because only the maximum of the $N$ bins is stored, there is a substantial probability that a time bin with a true signal whose amplitude is small compared to the noise fluctuation (i.e., when $F(E_{s}|0)< 1)$ will be unrecorded because there will be a larger noise/background fluctuation in one of the other $N-1$ bins. By contrast, for large amplitude signals, $F(E_{s}|0)\approx 1$, $f_\mathrm{max}(E'|E_{s}) \approx f(E'|E_{s})$. This matches our intuition: large merge windows will tend to suppress the probability that the trigger algorithm finds small true signals. 

Writing the measured differential rate (Eq. \ref{eq:dRdE'_Tmerge}) in terms of $f(E'|0)$ using Eq.~\ref{eq:fmax} and \ref{eq:fmax_Es} we find 
\begin{widetext}
\begin{equation}
\begin{split}
\frac{dR}{dE'}(E'|S) &= \frac{1}{\Delta t_\mathrm{OF}}f(E' | 0)F(E'|0)^{N-1} \\
                               &+ R_{s} F(E'|0)^{N-1}\left( {\left[ f(E'| E_{s}) \!-\! f(E'|0) \right]} \!+\! {\left[F(E'|E_{s}) \!-\! F(E'|0) 
\right] }  \frac{(N \!-\! 1) f(E'|0)}{F(E'|0)}  \right)
\end{split}
\label{eq:dRdE'_Tmerge2}
\end{equation}

Due to the facts that $F(E'|0))^{N-1} < 1$ and $[F(E'|E_{s}) - F(E'|0)] <0 $ always, the net differential sensitivity is always suppressed by a large merging window, though, of course, the sensitivity loss is most egregious for small signals. Consequently,  we strongly recommend one uses the smallest feasible merge window when searching for signals that comingle with noise. However, when searching for large signals where there is no overlap with the noise distribution as occurs in high mass dark matter searches when $F(E_{s}|0) \sim 1$, Eq.~\ref{eq:dRdE'_Tmerge} simplifies back to the expected 
\begin{equation}
\lim_{F(E_{s}|0)\rightarrow 1}\frac{dR}{dE'}(E'|S) = \frac{1}{\Delta t_\mathrm{OF}}f(E' | 0)+ R_{s} f(E'| E_{s})
\label{eq:dRdE'_Tmerge_linear}
\end{equation}
\clearpage
\end{widetext}

\section{Measuring Potentially Contaminated Net Differential Signals in Analog Detectors with Salting }
 \label{sec:salting}
In Sec.~\ref{sec:noDM}, we showed that conservative limits on signal interaction rates could be estimated from the \textbf{measurable} potentially signal contaminated background distribution $f(E'|S)$ and the measurable potentially contaminated net differential signal sensitivity $\Delta f(E'|S+s)=f(E'|S+s)- f(E'|S)$ for an idealized integrating detector in the no true signal pileup regime. Then in Sec.~\ref{sec:waveform_detector}, we showed that analog detectors with a continuous time stream were qualitatively similar to the idealized integrating detectors and thus all of our intuition remains intact and we should be able to follow a similar procedure to produce conservative signal rate limits.  Unfortunately, the quantitative detailed differences require a change in approach. Specifically:
\begin{itemize}
\item In analog detectors, the correlation time $\Delta t_\mathrm{OF}$ and even the concept of coincidence are qualitative concepts. Consequently, the simplistic relationship that $\frac{dR}{dE'}= \frac{1}{\Delta t} f(E')$ is not rigorously correct.  As a consequence, we will work directly with the differential rates such as $\frac{dR}{dE'}(E'|S)$.
\item  Triggering and merging algorithms are all fundamentally  non-linear. As a result, $\frac{dR}{dE'}(E'|S+E_{s})\neq \frac{dR}{dE'}(E'-E_{s}|S)$.  In Sec.~\ref{sec:waveform_detector}, we did produce a qualitatively accurate, but very simplistic model of these non-linear effects for a specific trigger algorithm but using this model for sensitivity estimates would certainly incur some systematic modeling error. 
\end{itemize}

What remains true for analog detectors is that the dependence of the search differential rate, $\frac{dR}{dE'}(E'|S)$, on the signal rate can be Taylor expanded to first order in the signal rate as long \textbf{as the true signal pileup rate is small}:
\begin{equation}
\begin{split}
   &\frac{dR}{dE'}(E'|S(\{R_{si},E_{si}\})) \\
   &=\frac{dR}{dE'}(E'|0) +\sum_{i}R_{si} \Delta f(E'|E_{si})+ ...
\end{split}
\label{eq:dRdE(E'|S)_linear-mul}
\end{equation} 
 Here, we have again made explicit that $S$ describes the potential superposition of signals of various true energies (the sum can also include integration over true energies). This expresses that, in the no-pileup limit, both the increase in measured energy differential rate and the dead time it imposes on the noise and background differential rate are both proportional to $R_{si}$. The constant of proportionality is as before, the net differential sensitivity for a signal of that true energy. We keep the same notation, $\Delta f(E'|E_{si})$, although it is no longer a simplistic, easily written difference between $f$'s. 

Following the same logical arguments that led to Eq.~\ref{eq:dRdE'_conservative_linear1} and ~\ref{eq:dRdE'int}, we show that salting the signal contaminated search data can be used to estimate $\Delta f(E'|E_{si})$ with an accuracy sufficient for approximation to first order in the rate used in Eq.~\ref {eq:dRdE(E'|S)_linear-mul}.

Specifically, let us add a known random rate $r_{s}$ of events with true energy $E_{s}$ \textbf{to the raw analog search data stream before the triggering algorithm}. To first order in rates, the salted events behave in the same way as the signal events $S_i$  and Eq.~\ref{eq:dRdE(E'|S)_linear-mul} becomes
\begin{equation}
\begin{split}
   &\frac{dR}{dE'}(E'|S(\{R_{si},E_{si}\})+s(r_s,E_s)) \\
   &=\frac{dR}{dE'}(E'|0) +\sum_{i}R_{si} \Delta f(E'|E_{si})\\
   &+ r_{s} \Delta f(E'|E_s)\\
   &=\frac{dR}{dE'}(E'|S(\{E_{si},R_{si}\}))
   + r_{s} \Delta f(E'|E_s)\\
\end{split}
\label{eq:dRdE(E'|S)_salting}
\end{equation} 
Therefore, to first order in signal rates the net differential sensitivity
\begin{equation}
   \Delta f(E'|E_s)=\frac{\frac{dR}{dE'}(E'|S+s)-\frac{dR}{dE'}(E'|S)}{r_s}
\label{eq:diffSens_salting}
\end{equation} 
 can be estimated from the salting rate normalized difference between the measured salted and unsalted differential rates: 
\begin{equation}
\widehat{\Delta f}(E'|S+E_{s}) = \frac{\widehat{\frac{dR}{dE'}}(E'|S+s)-\widehat{\frac{dR}{dE'}}(E'|S)}{r_{s}}
\label{eq:DeltaF_salt}
\end{equation}
This result makes intuitive sense. The difference between the post-analysis differential rates between the salted and non-salted searches is the net effect of the salted signal interaction. At small signal rates, the probability of pileup between salt and potential signals is negligible, and the theoretical and salted differential sensitivity are essentially the same.

This equation is valid for arbitrary $r_s$ under the limit of $r_s\Delta t_\mathrm{OF}\ll1$. The choice of $r_s$ does not affect the result, as it scales simultaneously in the denominator and numerator of Eq.~\ref{eq:DeltaF_salt}. It should not be confused with the potential DM signal rate $R_s$. In practice, a high $r_s$ may be desired for computational efficiency. As we have full control of the salting process, one can force the Monte Carlo process to avoid salt-salt pile-up. This procedure also avoids the subtlety of estimating $\Delta t_\mathrm{OF}$.

Notice that this salting scheme requires no definition of any {\it ad hoc} effective coincidence time scale. In fact, this scheme never attempts to explicitly define if the trigger is coincident or not coincident with the salted signal (which we know is a simplification since the weighting function, $\rho$, is a continuum and thus the concept of coincidence is also a continuum). Additionally, since the data is salted before the triggering algorithm, all subtleties, nonlinearities, and quirks of this algorithm are by construction accounted for, even if the algorithm is challenging to understand. 

If an estimate of the no-signal differential rate, $\frac{dR}{dE'}(E'|0)$ is possible, then background subtracted estimators for the signal interaction rate can be constructed using Eq.~\ref{eq:DeltaF_salt} to estimate the $\Delta f(E'|E_{s})$. 

If we do not have enough confidence to estimate no-signal differential rates, one can obtain in the no-pileup assumption a conservative upper limit for a putative signal in the following way.   We salt separately for each energy. This salting will be repeated on the raw data stream (not cumulatively) to estimate $\widehat{\Delta f} (E'|S+E_s)$ of each different $E_s$. Then, following Eq.~\ref{eq:dRdE'int}, we convolve the net differential response of each true energy with the true DM spectrum shape to estimate $\widehat{\Delta f}(E'|S+s(M_s))$. Finally, we set a conservative upper limit to the DM signal rate similarly as Eq.~\ref{eq:dRdE'_conservative}.
\begin{equation}
\widehat{R}_\mathrm{lim}\equiv \frac{\widehat{\frac{dR}{dE'}}(E'|S)}{\widehat{\Delta f}(E'|S+s(M_s))} 
\end{equation}

\section{Conclusion}

Previous light mass dark matter searches that set limits on dark matter interaction rates by searching for interactions whose sub-threshold true energy depositions were boosted above threshold by being in coincidence with large positive noise fluctuations did not account for the decreased non-signal affected live-time in their detector. To guarantee limit conservativeness while incorrectly modeling their experimental sensitivity, they added additional {\it post hoc} cuts that substantially decreased their sensitivity.

In this paper, we explicitly derive the net differential sensitivity estimator for idealized integrating detectors and show that it would give conservative interaction limits for all possible background scenarios, \textbf{provided that the true interaction signal rate produced minimal signal-signal pileup}. Furthermore, we propose two ways to ameliorate this limitation: first exclude the high signal pileup space with a conservative estimator\cite{das2024dark}, or restrict the limit-setting domain of the energy estimator to the non-Gaussian spectrum region only. 

We also showed that waveform detectors that produce a continuous time stream of data are qualitatively similar to idealized time discretized integrating detectors like CCDs in so far as there exists an effective timescale, $\Delta t_\mathrm{OF}$, below which noise is strongly correlated and signal interactions roughly add to the noise fluctuation, and above which noise is uncorrelated and signal interactions do not affect the noise and thus all of our intuition regarding estimating interaction rates in the linear regime remain valid, as long as one takes into account nonlinearities introduced by the triggering algorithm. This is most easily done by salting the raw data stream.

\section{Acknowledgments}
During the CPD analysis \cite{CPDv1_PRL21_LDMSearch}, Steve Yellin recognized and publicized internally within the SuperCDMS collaboration the fact that simply noise smearing the signal led to completely unphysical zero mass search sensitivity. This original insight (reproduced in Sec. \ref{sec:MDMto0}) directly led to and inspired this paper. Thus, even though he claims his contributions to this paper are minimal and aren't worthy of authorship, we strongly disagree. We also specifically recognize Wolfgang Rau for really pushing us to fully flesh out the ramifications of having a signal rate that had a significant pileup ($\lambda_{s}\geq1$). This push directly led to the realization that for pileup signal rates, the use of the linear net differential signal model could lead to non-conservative limits. We recognize and thank the SuperCDMS, TESSERACT, EDELWEISS, and RICOCHET collaborations for insightful comments throughout the drafting process, in particular Jules Gascon, Belina von Krosigk, Scott Oser, Wolfgang Rau, and Steve Yellin. This work was supported by the NSF and DOE.

\bibliography{reference} 

\end{document}